\documentclass[a4paper,11pt]{article}
\pdfoutput=1 %
\usepackage{jcappub} 

\usepackage[T1]{fontenc} 

\usepackage[utf8]{inputenc}

\usepackage{hyperref} 
\usepackage{url}
\usepackage{graphicx}
\usepackage{esvect}
\usepackage{graphics}
\usepackage{natbib}
\usepackage{mathrsfs}
\usepackage{blindtext}
\usepackage{comment} 
\usepackage{placeins} 

\usepackage[table]{xcolor}
\usepackage{multirow}

\usepackage[normalem]{ulem}
\usepackage{color}

\usepackage{soul}

\usepackage{supertabular}
\usepackage{tabularx}

\def\units#1{~\hbox{$\,{\rm #1}$}}

\title{Antiproton Bounds on Dark Matter Annihilation from a Combined Analysis Using the DRAGON2 Code}

\author[a, b]{Pedro~De~La~Torre~Luque}
\emailAdd{pedro.delatorre@uam.es}
\author[c]{Martin~Wolfgang~Winkler}
\emailAdd{martin.winkler@austin.utexas.edu}
\author[b]{Tim Linden}
\emailAdd{linden@fysik.su.se}

\affiliation[a ]{Instituto de Física Teórica, IFT UAM-CSIC, Departamento de Física Teórica,\\
Universidad Autónoma de Madrid, ES-28049 Madrid, Spain}
\affiliation[b ]{The Oskar Klein Centre, Department of Physics, Stockholm University, AlbaNova\\
  SE-10691 Stockholm, Sweden}
\affiliation[c ]{Department of Physics, The University of Texas at Austin, Austin, 78712 TX, USA}
\date{\today}

\abstract{
Early studies of the AMS-02 antiproton ratio identified a possible excess over the expected astrophysical background that could be fit by the annihilation of a weakly interacting massive particle (WIMP). However, recent efforts have shown that uncertainties in cosmic-ray propagation, the antiproton production cross-section, and correlated systematic uncertainties in the AMS-02 data, may combine to decrease or eliminate the significance of this feature. We produce an advanced analysis using the DRAGON2 code which, for the first time, simultaneously fits the antiproton ratio along with multiple secondary cosmic-ray flux measurements to constrain astrophysical and nuclear uncertainties. Compared to previous work, our analysis benefits from a combination of: (1) recently released AMS-02 antiproton data, (2) updated nuclear fragmentation cross-section fits, (3) a rigorous Bayesian parameter space scan that constrains cosmic-ray propagation parameters. 

We find no statistically significant preference for a dark matter signal and set strong constraints on WIMP annihilation to $b\bar{b}$, ruling out annihilation at the thermal cross-section for dark matter masses below $\sim200$~GeV. We do find a positive residual that is consistent with previous work, and can be explained by a $\sim70$~GeV WIMP annihilating below the thermal cross-section. However, our default analysis finds this excess to have a local significance of only 2.8$\sigma$, which is decreased to 1.8$\sigma$ when the look-elsewhere effect is taken into account.

}

\begin{document}
\maketitle
\flushbottom

\section{Introduction}
\label{sec:intro}

The Alpha Magnetic Spectrometer (AMS-02) has produced high-precision measurements of the local cosmic-ray antiproton spectrum, which has long been considered a critical channel for indirect dark matter (DM) searches~\cite{Giesen_2015, AMS_Ap}. Intriguingly, early studies in 2016~\cite{Cuoco, Cui2017} found statistically significant ($\sim5$$\sigma$) evidence for an excess over the expected astrophysical background in the cosmic-ray antiproton spectrum at an energy of $\sim$10~GeV that was best-fit by the annihilation of DM  particles with a mass of $\sim$60~GeV into $b\bar{b}$ final states.

Compared to the much-discussed cosmic-ray positron excess~\cite{Manconi:2020ipm, SERPICO20122, Hooper_2009} first observed by PAMELA~\cite{PAMELA:2008gwm}, and later by AMS-02~\cite{AMS02_positrons}, the DM  interpretation of the antiproton excess has a stronger theoretical motivation due to its consistency with a standard thermal WIMP. On the other hand, the suggested antiproton excess is more challenging to identify because the signal-to-background ratio is much smaller ($\lesssim$ 10\%). 

Due to the small relative amplitude of the excess, it is imperative to correctly account for systematic errors. Recent analyses have pointed out a number of effects that individually appear capable of decreasing, or entirely eliminating, the statistical significance of the excess, including: (1) uncertainties in cosmic-ray propagation and secondary production~\cite{Boudaud, Luque:2021joz, Korsmeier:2021brc, Weinrich:2020cmw, DelaTorreLuque:2020qml, Ap_Lin}, (2) uncertainties in the production cross sections of astrophysical antiprotons as well as other nuclear production cross-sections that are used as inputs for the cosmic-ray propagation codes~\cite{Reinert:2017aga,Korsmeier, Di_Mauro_Ap, delaTorreLuque:2022vhm, GenoliniRanking}, (3) correlations in the systematic uncertainties in the AMS-02 spectral data, which are undoubtedly present, but have not been publicly quantified by the AMS-02 collaboration~\cite{Boudaud,Heisig, Calore2021}.  While most studies agree that there is a positive residual in the antiproton flux at an energy of $\sim$10-20~GeV, the local significance of this excess spans from $\sim7\sigma$~\cite{Cholis} all the way down to 
a negligible $\sim1\sigma$~\cite{Heisig:2020jvs}. 


Recently, a new antiproton dataset was released by the AMS-02 collaboration, which included updated antiproton data spanning the period 2011-2018~\cite{AGUILAR20211}. A few recent studies have analyzed the antiproton spectrum using this dataset, generically finding that the size of the residual has decreased, and has also moved towards a higher antiproton energy. Thus, these analyses have found a statistically insignificant excess in the range of $0.5$$\sigma$ to $2$$\sigma$~\cite{Calore2021}, with best-fit DM  masses in the range of $80$~GeV to $120$~GeV.

In this paper, we 
utilize the DRAGON2 code in order to fit the new antiproton measurements from AMS-02, 
employing an advanced Bayesian analysis that constrains our fit to the antiproton data through simultaneous fits to the secondary spectra of B, Be, and Li. We include the halo height as a free parameter in our fit, along with the other propagation parameters and nuisance parameters that allow us to account for cross sections uncertainties. This comprehensive analysis improves on previous work due to the number of constraints (datasets) employed, the use of refined cross section parametrizations, the treatment of possible correlations in AMS-02 errors and the energy dependence of antiproton cross-section uncertainties.


We remark that there is no previous analysis which fits the different spectra (and ratios to primary particles, such as C and O) of B, Be and Li simultaneously with antiprotons. Recently,  efforts in this direction were made in Ref.~\cite{Calore2021}, where the authors map the probability distribution of the halo height -- obtained from a previous analysis of Li/C, Be/B and B/C -- into their antiproton fit, however keeping the remaining propagation parameters fixed. Furthermore, Ref.~\cite{Calore2021} is based on a slightly simplified semi-analytic propagation model as opposed to the fully numerical approach implemented in DRAGON2 and we make use of Bayesian inference.   

Focusing on DM  annihilation to $b\bar{b}$ final states, we find no statistically significant evidence for a DM  annihilation component. The largest local significance for an excess stands at $2.8\sigma$ for a DM  mass of $\sim 66$~GeV. On the one hand, this mass range is consistent with an interesting region of DM  models that are also capable of explaining the Galactic Center Excess (GCE)~\cite{Ackermann_2017}. On the other hand, the corresponding global significance the corresponding global significance, which takes into account the look-elsewhere effect, is only $1.8\sigma$. Utilizing these results, we place strong constraints on the DM  annihilation cross-section.

In Section~\ref{sec:Method}, we describe our simulation set-up and the methodology we employ in this work. We also compare how our set-up and analysis differs from other recent works in Section~\ref{sec:Comparison}. Then, in Section~\ref{sec:Sec_Ap} we report our fits to the antiproton spectra for the scenario where we only consider the production of antiprotons from CR interactions. We discuss and show our results for the combined analyses including DM antiproton production in Section~\ref{sec:DM_Ap}, where we also derive DM bounds.
Finally, we summarise and discuss our main findings in Section~\ref{sec:conc}.

\section{Methodology}
\label{sec:Method}

In this section we summarise the main points of the simulation procedure and analysis setup. We note that our procedure is close to that described in Ref.~\cite{Luque:2021ddh}, and we refer the reader to that paper for further technical details.

\subsection{Simulations setup}
\label{sec:Simsetup}

{\tt DRAGON2}~\footnote{\label{note1}\url{https://github.com/cosmicrays/DRAGON2-Beta\_version}} is an advanced CR propagation code designed to self-consistently solve the diffusion-advection-loss equation describing CR transport for all species involved in the CR network, including cosmic rays (CRs) of both astrophysical and exotic origin (e.g.,  from DM annihilations/decays). The transport equation features fully position-dependent and energy-dependent transport coefficients in either two-dimensional (assuming cylindrical symmetry) or three-dimensional configurations of the Galaxy structure (i.e. considering the spiral arm pattern of the Galaxy). The flexibility of the {\tt DRAGON2} code allows for detailed studies of both small-scale and large-scale structures (e.g., the spiral structure of the Galaxy) in steady-state and transient mode, refining the spatial resolution on the regions of interest (e.g., local bubble, GC, or Galactic Plane).

In this work, we solve the coupled system of propagation equations for all the isotopes involved in the CR network~\cite{Tomassetti:2017hbe} up to $Z=14$ (Silicon) with a customised version of the {\tt DRAGON2} code~\cite{DRAGON2-1, DRAGON2-2} that is optimized for studies of antiprotons and antinuclei~\footnote{\label{note2} Publicly available at \url{https://github.com/tospines/Customised-DRAGON2_Antinuclei}}.
For these analyses, we solve these equations assuming cylindrical symmetry in the Galaxy, which makes use of a homogeneous gas distribution implemented according to Ref.~\cite{ferriere2007spatial}. We describe the distribution of CR sources (expected to be mainly supernova remnants - SNRs)~\cite{bell1978acceleration, blandford1978particle, axford1982structure} using the spatial distribution derived by Ref.~\cite{Lorimer_2006}. We parameterise the source injection as a broken-power law in energy per nucleon.

We set the diffusion coefficient to be spatially independent, with a rigidity dependence described by a broken power-law modeled as:
\begin{equation}
 D (R) = D_0 \beta^{\eta}\frac{\left(R/R_0 \right)^{\delta}}{\left[1 + \left(R/R_b\right)^{\Delta \delta / s}\right]^s} ,
\label{eq:diff_eq}
\end{equation}
where we fix $R_0 = 4\units{GV}$, the value of the rigidity break $R_b = 312 \units{GV}$, the change in the spectral index $\Delta\delta = \delta - \delta_h $ = 0.14, and the smoothing parameter $s = 0.040 $, following the values determined in Ref.~\cite{genolini2017indications} from a detailed analysis of the proton and helium AMS-02 fluxes.

We consider reacceleration, which is directly related to the spatial diffusion coefficient, and set the Alfv{\'e}n velocity as a free parameter in our study. However, we neglect convection, since analyses of secondary-to-primary ratios at AMS-02 
do not favor any relevant impact of convection on CR propagation~\cite{Weinrich:2020cmw, Niu:2017qfv, Niu:2018waj, Luque:2021nxb}.

We calculate inelastic cross-sections using the CROSEC parameterisation~\cite{Barashenkov:1994cp}, but utilize the DRAGON2 inclusive spallation cross-sections for the production of secondary nuclei ($Z>1$)~\cite{Luque:2021joz, Evoli:2019wwu}. We note that there have been significant efforts over the last few years dedicated to improving the current cross section parameterisations and to account for their uncertainties in the analysis. Other commonly considered parameterizations include those from GALPROP~\cite{GALPROPXS, GALPROPXS1} or FLUKA~\cite{delaTorreLuque:2022vhm} (see also the recent work by~\cite{Maurin_2022}). 
However, the DRAGON2 cross sections have recently been demonstrated to provide a better simultaneous fit to the B, Be and Li flux ratios, compared to other cross sections data-sets~\cite{Luque:2021nxb}.
For a comparison between different cross section models for some of the main channels involving the production of B, Be and Li, we refer the reader to Appendix~\ref{sec:appendixA}.

We utilize cross sections for antiproton production from CR proton and helium collisions following the calculations in Ref.~\cite{Winkler:2017xor} (''Winkler`` option in the {\tt DRAGON2} code).
In addition, we take into account the tertiary contribution for production of protons and antiprotons.
In this work, we also account for the production of $\bar{p}$ particles from nuclei heavier than He. This is done by scaling the proton spallation cross sections by a factor of $A^{0.9}$~\cite{Korsmeier}, where $A$ is the mass number of the CR nucleus colliding with ISM gas. This results in a slight increase ($\sim$3\%) in the $\bar{p}$ production rate with respect to our previous work~\cite{Luque:2021ddh}.

Additionally, in our analysis, we do not consider the $\bar{p}/$p data below $4$~GV, in order to limit the effect of the uncertainties in solar modulation in our results. This cut leads to more conservative limits for low DM masses (below $\sim100$~GeV). In our calculations we implement solar modulation using a modified Force-Field approximation~\cite{forcefield} which accounts for the effects of charge-sign dependence~\cite{Cholis:2015gna}. In this case, the Fisk potential takes the following form:
\begin{equation}
\phi^{\pm} (t, R) = \phi_0(t) + \phi_1^{\pm}(t) F(R/R_0) ,
\label{eq:Charge-sign_Modul}
\end{equation}
parameterising its rigidity-dependence as $F(R/R_0) \equiv \frac{R_0}{R}$, with $R_0 = 1 \units{GV}$~\cite{Reinert:2017aga}. 

The $\phi_1^+$ ($\phi_1^-$) term accounts for the additional energy loss of positively (negatively) charged particles during a negative (positive) solar polarity phase. This additional energy loss occurs because particles with charge opposite to the magnetic polarity access the heliosphere by complicated inward drift along the heliospheric current sheet, while particles whose charge aligns with the polarity can enter on direct trajectories along the poles. Since AMS-02 data were mostly taken during a positive solar polarity phase we set $\phi^+_{1,AMS-02} = 0 $ and only consider a non-vanishing $\phi^-_{1,AMS-02} $. 
Neglecting the short time-period of opposite polarity -- the polarity flip occurred near the beginning of AMS-02 operation (around 2012) -- should not affect our analysis, in particular as we only include the $\bar{p}/$p data above $4$~GV.

A reasonable value of the constant Fisk potential, $\phi_0$, was found to be $0.61 \units{GV}$ for the period of 2011-2016 (relevant for the AMS-02 data-sets for the secondary-to-primary ratios, C, O and He)~\cite{Aguilar:2018keu}, since it allows us to reproduce Voyager-1~\cite{stone2013voyager,cummings2016galactic} data and AMS-02 data in that period, while also remaining consistent with the NEWK neutron monitor experiment\footnote{Data and information available at \url{http://www01.nmdb.eu/station/newk/}} \cite{MaurinCRDB, MAurinCRDB2} (see also~\cite{ghelfi2016non,ghelfi2017neutron}). 
In order to adjust the value of $\phi_0$ to the data taking period 2011-2018 (relevant for the newly released antiproton data-set~\cite{AGUILAR20211} as well as for the primary CRs Ne, Mg, Si~\cite{Aguilar:2020ohx} and the most recent H and He data sets) we consider the time-variation in the neutron intensity between 2011 and 2018 as detected by NEWK~\cite{ghelfi2016non, ghelfi2017neutron, MaurinCRDB, MAurinCRDB2}. In this way we obtain $\phi_0 = 0.58 \units{GV}$ for the 2011-2018-period.
Because charge-dependent effects (mainly linked to propagation within the heliospheric current sheet) are necessary to explain the different behaviour of electrons and positrons~\cite{AMS-02_TimeSeries}, we allow that negative CR particles can have a non-vanishing $\phi^+_{1}$, finding a good agreement with the data for antiprotons using $\phi^-_{1}$ around $0.9 \units{GV}$, consistent with Ref.~\cite{Reinert:2017aga, Heisig, Luque:2021ddh}. 

We stick to the more conventional setup for our analysis, although we note that we neglect other possible astrophysical processes that could significantly affect the propagation of CRs and their production. For example, we neglect the spatial dependence of the diffusion coefficient, while there are both theoretical arguments and $\gamma$-ray observations of the diffuse emission (see, e.g.~\cite{delaTorreLuque:2022vhm, Lipari, Cerri2017jcap, LHAASO_Diff}) supporting a change of the diffusion coefficient in different zones of the Galaxy. It was shown that models that consider possible differences in the transport of CRs in the galactic halo and disk predict only a slightly different antiproton spectrum measured at Earth~\cite{Zhao:2021yzf, Feng2016}.
Additionally, the production of secondary CRs within the acceleration region of supernova remnant can change the spectra of secondary CRs~\cite{Mertsch:2014poa, Kohri_Ap, Sec_SNRs, Cholis_ApSNR}. 
These modifications of the standard scenario of Galactic CR transport could affect the astrophysical searches for DM  signatures and will be explored in a future work.

\subsection{Analysis setup}
\label{sec:Ansetup}

As in our previous work~\cite{Luque:2021ddh}, our set-up (similar to the one tested in Ref.~\cite{Luque:2021nxb}, where more details are given) relies on a Markov chain Monte Carlo (MCMC) analysis that is based on Bayesian inference~\cite{emcee}. We determine the propagation parameters involved in the transport equation (namely H, $D_0$, $V_A$, $\eta$ and $\delta$) from a combined fit to AMS-02 spectral data for the main secondary CR nuclei (B, Be and Li) along with the $\bar{p}$ spectra. Many previous studies~\cite{GenoliniRanking, Weinrich:2020cmw, Luque:2021joz, delaTorreLuque:2022vhm, Korsmeier:2021brc} have shown the importance of combining different secondary CR observations in order to mitigate the effect of systematic uncertainties (primarily those associated with the production cross-sections of secondary CR) on the propagation parameters of the simulation. Our fitting procedure includes the injection parameters of the primary CRs included in our simulation set-up ($^1$H, $^{4}$He, $^{12}$C, $^{14}$N, $^{16}$O, $^{20}$Ne, $^{24}$Mg and $^{28}$Si).
In this work, we fit the B/C, B/O, Be/C, Be/O, $\bar{p}/$p flux ratios and the $^{10}$Be/Be and $^{10}$Be/$^{9}$Be flux ratios (allowing us to constrain the halo height, H~\cite{Donato:2001eq}), as well as the Li/B flux ratio. We neglect correlations between the nuclear CR data sets (for instance between B/C and B/O) and treat them as statistically independent. This is justified since such correlations only slightly affect the uncertainties in the propagation parameters, and have almost no impact on the DM bounds we compute in this work. 

The Fisk potential is adjusted for the period of data collection (2011-2018), as described above.
Moreover, in this analysis we include nuisance parameters that allow us to modify the normalization of the spallation cross sections, since spallation cross uncertainties above a few GeV/n mostly affect the overall normalization~\cite{Korsmeier, Cuoco2, Di_Mauro_Ap, Luque:2021joz}. Extending the analysis of our previous work, a scale factor, that allows us to apply a constant offset in the production rate of antiprotons, has also been added as a nuisance parameter for the $\bar{p}$ cross sections, so that there are four total nuisance parameters (scale factors) included in this analysis: $\mathcal{S}_{B}$, $\mathcal{S}_{Be}$, $\mathcal{S}_{Li}$ and $\mathcal{S}_{Ap}$. These factors allow the model to further adjust the production of these secondary nuclei and are necessary to alleviate the tension between the grammage inferred from analyses of secondary-to-primary ratios of secondary CR nuclei (B, Be and Li) and the grammage predicted by analyses of the antiproton-over-proton flux ratio, which reflects the impact of cross sections uncertainties in the production of these secondary particles, as discussed in Ref.~\cite{Luque:2021ddh}. 

An important point of this analysis is that we include, in the likelihood definition, a penalty factor associated to each of our nuisance scale factors (see more details in Ref.~\cite{Luque:2021nxb}). In practice, this approach penalizes large variations from the original cross sections, and can be interpreted as an implementation of existing constraints on production cross-sections of B, Be, Li and $\bar{p}$. These penalty factors are inversely proportional to the relative variance of the cross section data for the channels that we consider (similar to what is defined in Eq. 4 of Ref.~\cite{Weinrich:2020cmw}). In this case, for B, Be and Li we consider the variance in the cross section data for the channels of production of each of their isotopes ($^{11}$B, $^{10}$B, $^{10}$Be, $^{9}$Be, $^{7}$Be, $^{7}$Li, $^{6}$Li) with C and O colliding with p~\cite{Luque:2021joz}.

For the $\bar{p}$ production cross sections, the amount of experimental data is more limited, and we only include the variance for the data of direct $\bar{p}$ production from p+p collisions \footnote{The average (relative) standard deviation for B, Be and Li cross sections data from $^{12}$C-p ($^{16}$O-p) interaction is $18.6\%$ ($24\%$), $14.7\%$ ($26.9\%$), $21.2\%$ ($23\%$), respectively~\cite{Luque:2021joz}, while for the production of antiprotons in p-p collisions we find it to be around $12\%$ (see Refs.~\cite{Winkler:2017xor, Korsmeier}).}. 
Given the difficulty of assessing the systematic uncertainties related to antiproton production we adopt two different analyses that differ in how we treat antiproton cross sections: A simplified analysis, where there is only a scale factor for the antiproton cross sections and its penalty factor has an associated variance of $12\%$ (we call this analysis the ``Simplified'' analysis) and a more advanced analysis based on Ref.~\cite{WinklerDiMauro} where we include a covariance matrix for the energy dependence of antiproton cross section uncertainties and the penalty associated with the scale factor has a variance of $6.6\%$. We call this the ``Canonical'' analysis. In both cases, scaling factors for the cross sections of CR secondary nuclei (namely B, Be and Li) are included.

For both analyses, we study scenarios where we include the contribution of a WIMP annihilating into $b\bar{b}$ final states that subsequently produces $\bar{p}$. We then test whether a DM  contribution is preferred and derive constraints on the DM  mass and annihilation cross section ($\left< \sigma v \right>$) along with the rest of transport parameters and nuisance factors. The annihilation yield at production (dN/dE) for every DM  mass is computed using the PPPC4DM package~\cite{Marco_Cirelli_2011, Ciafaloni_2011}. 
We use the Navarro-Frenk-White (NFW) DM density profile~\cite{Navarro:1995iw} as a reference with a DM  density of $\rho_{\odot} = 0.4\units{GeV}$ at the Solar System position, $r_{\odot} = 8.3\units{kpc}$~\cite{Fabio_Iocco_2011}.

In our procedure, we incorporate correlations in the AMS-02 systematic errors through covariance matrices computed in Ref.~\cite{Heisig}, which primarily affects the statistical significance of the reported anomaly, and only slightly changes the diffusion parameters obtained in our fits and the (best-fit) predicted spectra of secondary CRs. We remind the reader that these matrices are not released by the AMS-02 collaboration and only constitute best estimates obtained with the publicly available information. 
Therefore, our main tests will consist of the Canonical and Simplified analyses considering correlations in AMS-02 systematic errors and, for comparison, we also present the results obtained assuming no correlation in AMS-02 errors.

In order to reduce the impact of uncertainties in solar modulation, we do not consider in the analysis the $\bar{p}$ ratios below $4\units{GeV}$, as explained above. As tested in our previous work, analyses that use alternative energy cuts at $3\units{GeV}$ or $5\units{GeV}$ do not significantly change the results. We also test that small variations on the parameter $\phi^-_{1}$ do not affect our results. This confirms the robustness of our analysis with respect to solar modulation uncertainties.

The experimental data that we employed in this work~\cite{Aguilar:2015ooa, aguilar2018observation, AGUILAR20211, ACEBe, IMP1, IMP2, ISEE, Hams_2004, UlysesBe, VoyagerMO} were taken from the ASI Cosmic Ray Data Base~\cite{Pizzolotto:2017lfd}~\footnote{\url{https://tools.ssdc.asi.it/CosmicRays/}} and the Cosmic-Ray DataBase~\cite{2020Univ....6..102M}~\footnote{\url{https://lpsc.in2p3.fr/crdb/}}.

\subsection{Comparison with other recent analyses of the 2018 $\bar{p}$ data-set}
\label{sec:Comparison}

In addition to our previous work, several other groups have analyzed the new AMS-02 $\bar{p}$ data set over the last two years~\cite{Kahlhoefer2021, Calore2021, WinklerDiMauro, Ding_2023, Balan:2023lwg}. We briefly review the analysis choices of each of these studies here:

In the analyses of Kahlhoefer et al and Balan et al (Refs.~\cite{Kahlhoefer2021} and~\cite{Balan:2023lwg}), the authors derive bounds on the DM annihilation rate by a joined fit to the hydrogen and helium spectra and the $\bar{p}/$p ratio. The focus of these works is, however, on the implementation of machine learning techniques into CR analysis, rather than on a fully realistic modeling of all relevant systematics (which would require, for instance, the inclusion of additional CR species to constrain the halo height). They show how different diffusion setups (i.e. different ingredients, such as convection or reacceleration, and assumptions on the injection and breaks in the diffusion coefficient) lead to significantly different bounds.

In the analysis of Di Mauro and Winkler~\cite{WinklerDiMauro}, they derived bounds on DM  annihilation from a joined fit to the AMS-02 antiproton and B/C data. While the modeling of cross sections and systematic uncertainties is similar to our work, Ref.~\cite{WinklerDiMauro} considered a few fixed choices of the halo height rather than deriving it within the fit (which would require the inclusion of radiactive species as in our work). Furthermore, Ref.~\cite{WinklerDiMauro} employs the semi-analytic two-zone diffusion model for CR propagation as opposed to our fully numerical approach with DRAGON2.
In Calore et al.~\cite{Calore2021}, the authors derive DM limits by a fit to the AMS-02 antiproton spectrum. Systematic uncertainties in the antiproton cross sections, as well as AMS-02 error correlations are included through covariance matrices similar to Ref.~\cite{WinklerDiMauro} and our work. In order to include the halo height as a fit parameter the authors extract its probability distribution from an earlier CR study on secondary-to-primary ratios. Uncertainties on the remaining propagation parameters are included in a simplified way through a covariance matrix. In contrast to our work, Ref.~\cite{Calore2021} employs the two-zone diffusion model for CR propagation. This analysis is mainly based on the 2015 $\bar{p}$ data-set, while we analyze the 2018 dataset, which includes nearly a factor of two improvement in the exposure.

More recently, Ding et al.~\cite{Ding_2023} derived DM  bounds from a combined fit to AMS-02 protons and antiprotons (2011-2018 dataset), using the {\tt GALPROP} code, where the propagation parameters were fixed to the ones that reproduce the B/C ratio. The antiproton cross sections are those from the {\tt GALPROP} v.54 code and neither cross section uncertainties nor other systematic errors are taken into account.


Our study contains a number of improvements compared to these earlier works. In particular, we perform the first combined fit to the AMS-02 $\bar{p}/p$ data and a large set of secondary-to-primary and secondary-to-secondary CR ratios. In this way we are able to fully include all relevant propagation uncertainties in our analysis. Specifically, we directly constrain the halo height $H$, which is a key parameter in dark-matter studies, within our fit. This is in contrast to previous works which, at best, considered some fixed values of $H$ (with the exception of Ref.~\cite{Calore2021} which incorporated a simplified probability distribution for $H$). Another new ingredient are the improved DRAGON2 spallation cross sections which resolve some tensions in the fits to nuclear cosmic-ray spectra and allow for a more accurate determination of the CR propagation parameters. Furthermore, we perform a state-of-the art modeling of the relevant cross section uncertainties and of the error correlations in the AMS-02 data -- which is missing in some of the previous works. In this light, our analysis represents the most advanced study of the AMS-02 2018 antiproton data set performed so far.

\section{Results of the combined antiproton analyses}
\label{sec:results}


\subsection{Secondary antiproton spectrum}
\label{sec:Sec_Ap}

\begin{figure}[!t]
\centering
\includegraphics[width=0.47\textwidth, height=0.22\textheight]{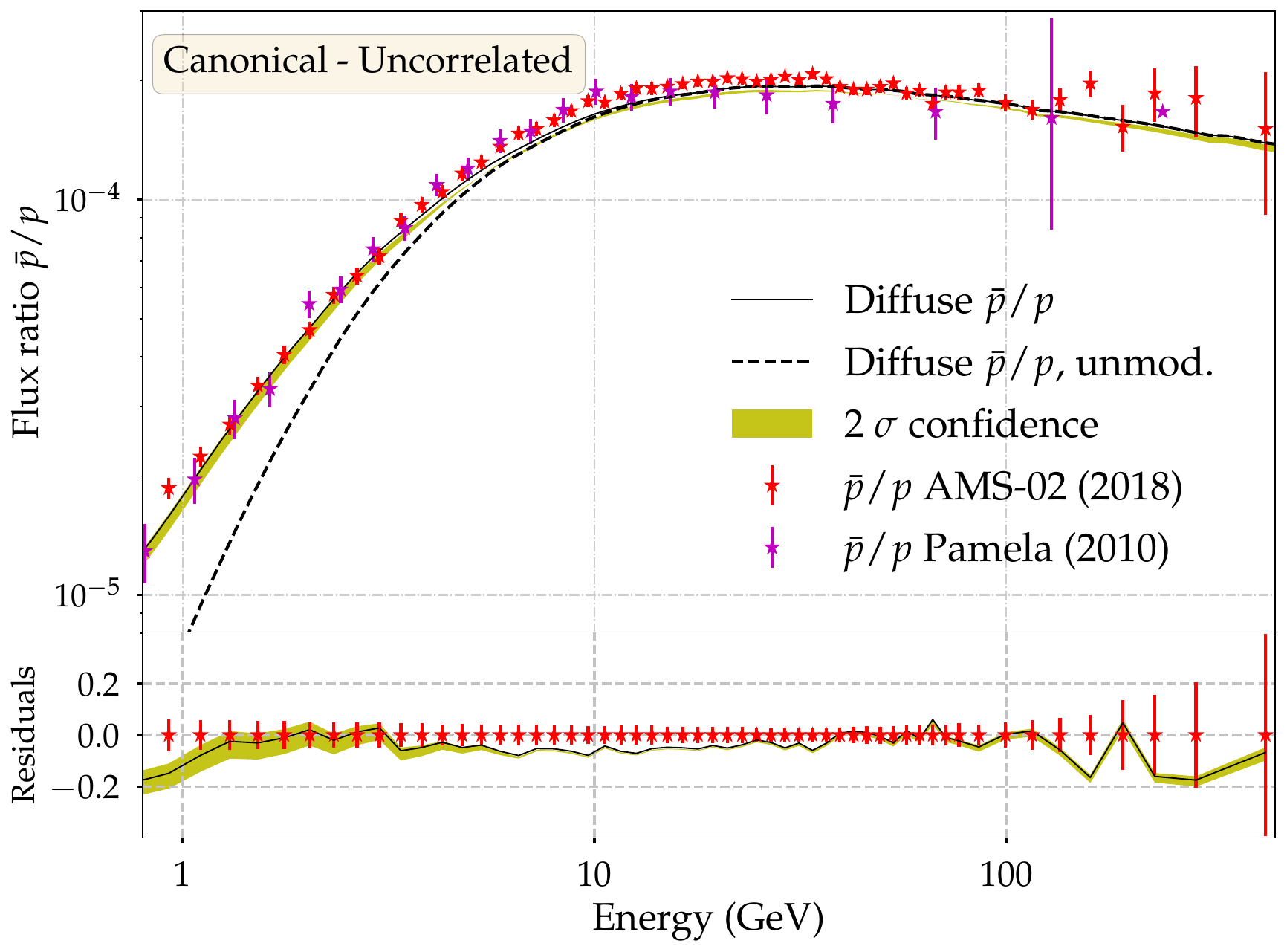} 
\includegraphics[width=0.52\textwidth, height=0.24\textheight]{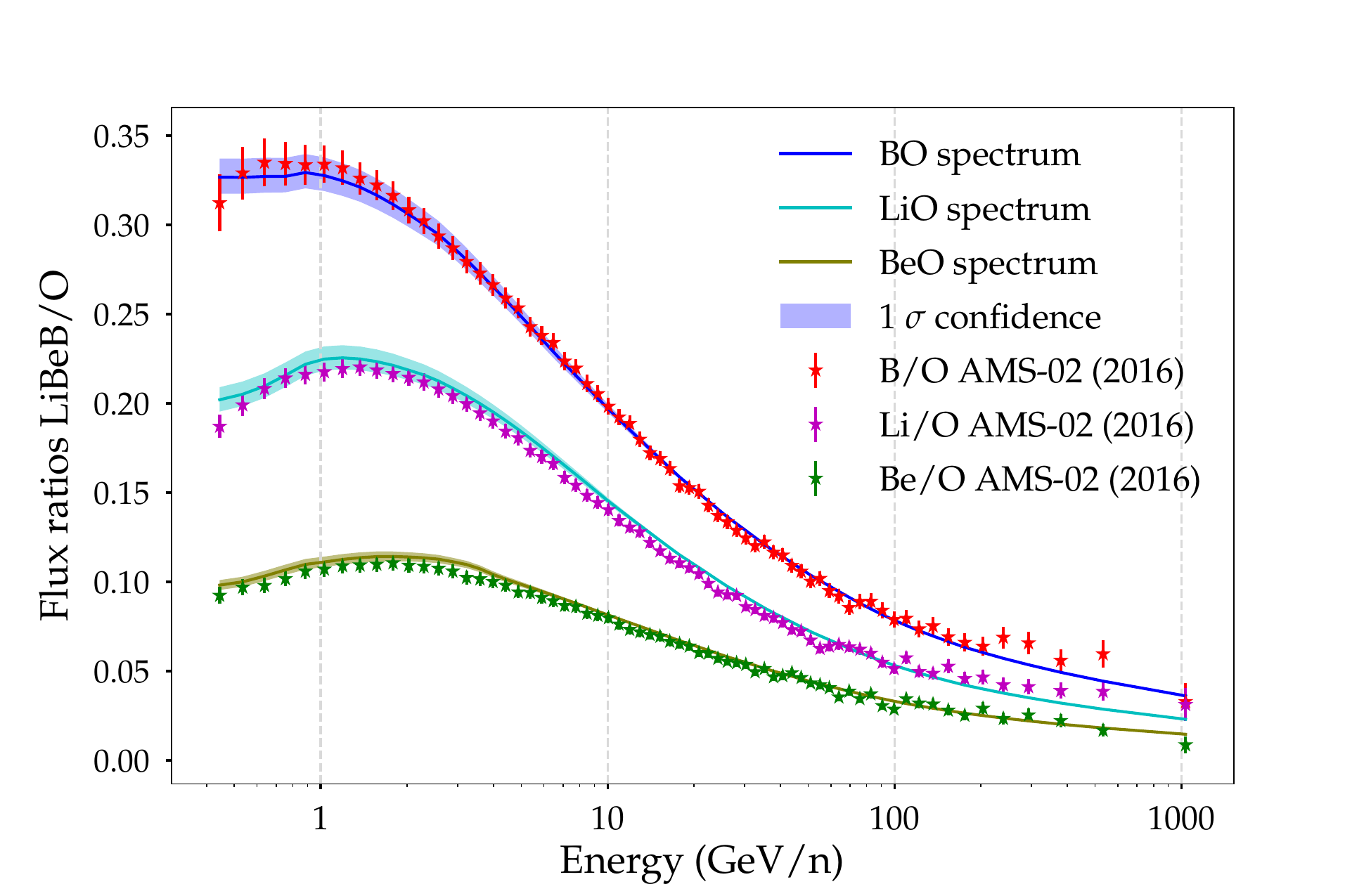}

\includegraphics[width=0.47\textwidth, height=0.22\textheight]{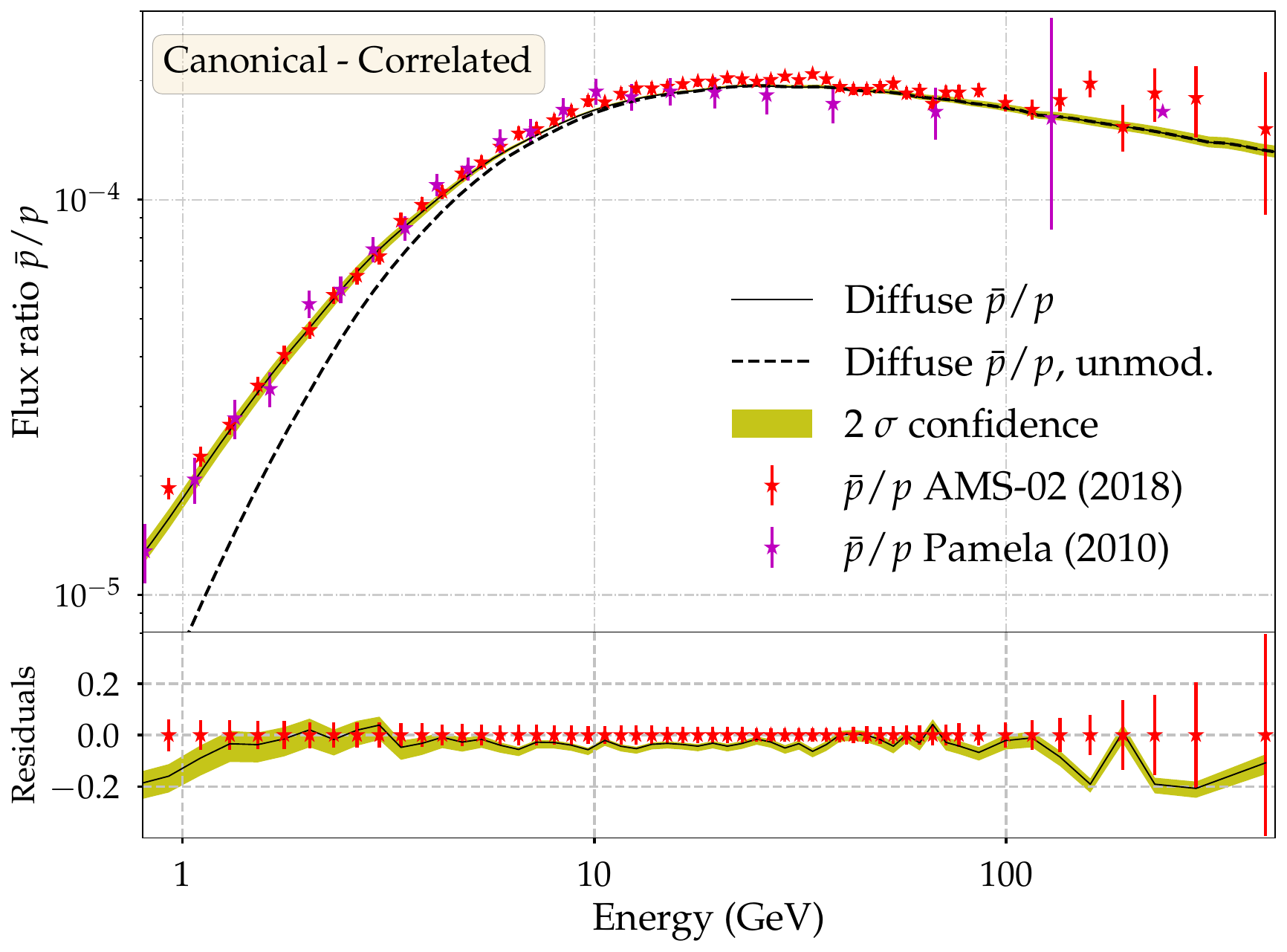} 
\includegraphics[width=0.52\textwidth, height=0.24\textheight]{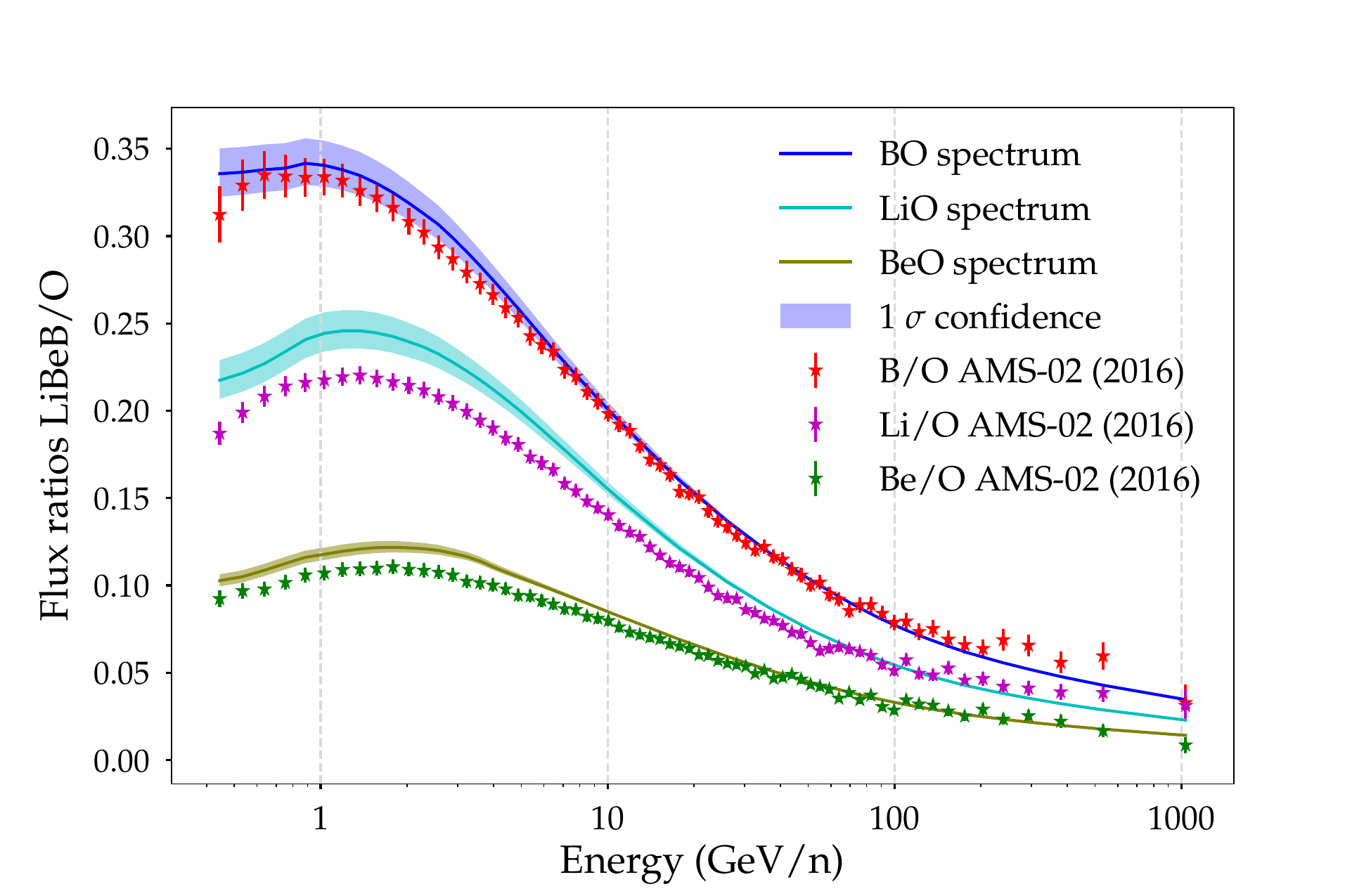}
\caption{ $\bar{p}/p$ spectrum (left panels) and secondary-to-primary flux ratios of B/O, Be/O and Li/O (right panels) evaluated with the transport parameters obtained in our Canonical analyses in a scenario that does not include any DM production of antiprotons. The top row shows the spectra obtained from the analysis including no correlation in the AMS-02 systematic errors while the bottom row shows the spectra for the case when we include correlations. The variation in the flux ratios due to the uncertainty in the determination of the propagation parameters is shown (not including modulation uncertainties) in each case and the error bars on AMS-02 data are the $1\sigma$ errors reported by the collaboration.  The residuals in the left panels are drawn with respect to AMS-02 data (2018).
}
\label{fig:App_NoWimp}
\end{figure}

Here, we report and compare the results from the combined antiproton analyses without including DM contributions to the antiproton spectrum.
In Figure~\ref{fig:App_NoWimp}, we show the spectra of the main secondary flux ratios ($\bar{p}/$p flux ratio, in the left column, and ratios of B/O, Be/O and Li/O, in the right column) derived from the best-fit propagation parameters~\footnote{In this paper, the best-fit parameters are determined from the probability distribution function obtained for every parameter in the MCMC procedure described above. The $1\sigma$ uncertainty bands reported are obtained by varying the diffusion parameters from the best-fit ones at the $68\%$ confidence level in their probability distribution function.} in the scenario of antiproton production solely from CR interactions. In particular, we show the results obtained from the Canonical analysis (where we include the covariance matrix for antiproton cross sections uncertainties) in the case where AMS-02 errors are treated as uncorrelated (top panels) and correlated (bottom panels).
The spectra obtained assuming correlated AMS-02 errors from the Canonical analysis and the Simplified one (where we only include a cross sections scaling factor) yield roughly the same spectral fits, although with the Simplified analysis we obtain larger cross sections scale factors.

In the analyses where we assume that the AMS-02 data errors are correlated, the quality of the fit is either $\chi^2=476.1$ (in the case where we only include scaling factors for cross sections, Simplified analysis), or $\chi^2=553.4$ (in the case where the cross sections covariance matrix is included, Canonical analysis). In both cases, there are $329$ degrees of freedom. Meanwhile, we obtain a $\chi^2\sim300$ in the uncorrelated cases.
The low $\chi^2$/d.o.f obtained from the uncorrelated case points to the fact that we are overestimating the uncertainties in AMS-02 data if we assume that errors are uncorrelated~\cite{Boudaud, derome2019fitting}. Indeed the AMS-02 measurements are dominated by systematic uncertainties that have been shown to be highly correlated~\cite{Heisig}. However, given the difficulties on estimating all the factors accounting for these systematic errors in the AMS-02 detector, our modifications to the $\chi^2$ can only be thought of as estimates based on publicly available information.

The probability distribution functions (PDFs) obtained for every parameter are shown in Appendix~\ref{sec:diff_params}. We find very flat residuals in the $\bar{p}/$p ratio with respect to the 2018 data-set, such that the $\bar{p}$ spectra are easily reproduced with a constant scaling of their production cross sections. This result is similar to both our earlier work and that of Ref.~\cite{WinklerDiMauro}. For the Simplified analysis, we find a scaling factor of $\mathcal{S}_{Ap} = 1.11\pm0.02$ (a $\sim11\%$ increase in the cross sections of $\bar{p}$ production) and $\mathcal{S}_{Ap} = 1.14^{+0.01}_{-0.02}$ for the uncorrelated and correlated cases, respectively. We highlight that these scale factors are compatible with the variance associated to the $\bar{p}$ cross sections data in this analysis (which is $\sim$$12\%$~\cite{Korsmeier}) and are similar to those obtained in the analyses by Ref.~\cite{dimauro2023datadriven}.  
In our Canonical analysis (where we include the covariance matrix for the antiproton cross sections), we find a best-fit scaling of $\mathcal{S}_{Ap} = 1.07\pm0.01$ for the correlated case, which is still compatible with the $1\sigma$ cross-section uncertainties reported by Ref.~\cite{WinklerDiMauro} and a $\mathcal{S}_{Ap} = 1.08\pm0.01$ for the uncorrelated case.
The scaling associated with Be and Li is found to be within $8\%$, indicating that the cross sections of B production are compatible with very small or no scaling in both kind of analyses. 

The PDFs of the inferred propagation parameters in the scenario without DM  annihilation are shown in Figure~\ref{fig:joint_PDFs}, left panel, where we overlap the PDFs obtained in each of the analyses, including correlations in AMS-02 data. The transport parameters found in both analyses are compatible within $1\sigma$, which means that the determination of the propagation parameters is not significantly affected by the cross sections scaling model.

The halo height inferred in these analyses is $H~\sim4$~kpc and the ratio $D_0/ H$ is close to $10^{28}$~cm$^{2}$~s$^{-1}$~kpc$^{-1}$. The effective Alfv\`en velocity is in the range $V_A\sim 10-17$~km/s in the uncorrelated analyses and $V_A\sim 15-25$~km/s in the analysis that includes correlations in the uncertainties of the cosmic-ray data. 
Finally, the spectral index of the diffusion coefficient ranges from $0.45$ to $0.48$, compatible with a Kraichnan-like~\cite{kraichnan_1959, Kraichnan_review} turbulence spectrum of the plasma fluctuations below the break ($\sim 300$~GV) and Kolmogorov-like~\cite{Kolmogorov} above. This result is in good agreement other recent studies~\cite{Boschini2020, Korsmeier:2021brc, Weinrich:2020cmw}. 

These results confirm our earlier findings that current diffusion models allow us to simultaneously reproduce all the light secondary CRs (see also Ref.~\cite{delaTorreLuque:2022vhm}) within the context of a model that assumes only secondary production in the interstellar medium and a constant diffusion coefficient throughout the Galaxy. We note that if we decreased the uncertainties on the cross sections of production of secondary CRs, our results could clearly show tensions between the parameters needed to reproduce the different secondary CRs, therefore, future accelerator experiments could provide measurements that would help us to spot tensions and reveal signatures of new physics in astrophysical data.
A summary of the propagation parameters inferred is also given, in form of box plots, in Figure~\ref{fig:Boxplot_SecAp} and several examples of the best-fit {\tt DRAGON2} input files are available at \url{github.com/tospines/Customised-DRAGON-versions/tree/main/Custom_DRAGON2_v2-Antinuclei}. 

Finally, we remark that the fact that the energy dependence of the different secondary ratios agrees so well indicates that our simple diffusion scenario does not require significant modifications. However in order to further improve the search for new physics signatures, it will be necessary to also explore the implications of inhomogeneous diffusion in the Galaxy (which is suggested by some observations~\cite{Lipari, Luque:2022buq, Frontiers} and supported by theory~\cite{Cerri2017jcap, Feng2016}). We leave this analysis for future work. 

\subsection{Search for WIMP features in the antiproton spectrum}
\label{sec:DM_Ap}

\begin{figure}[!t]
\centering
\includegraphics[width=0.485\textwidth] {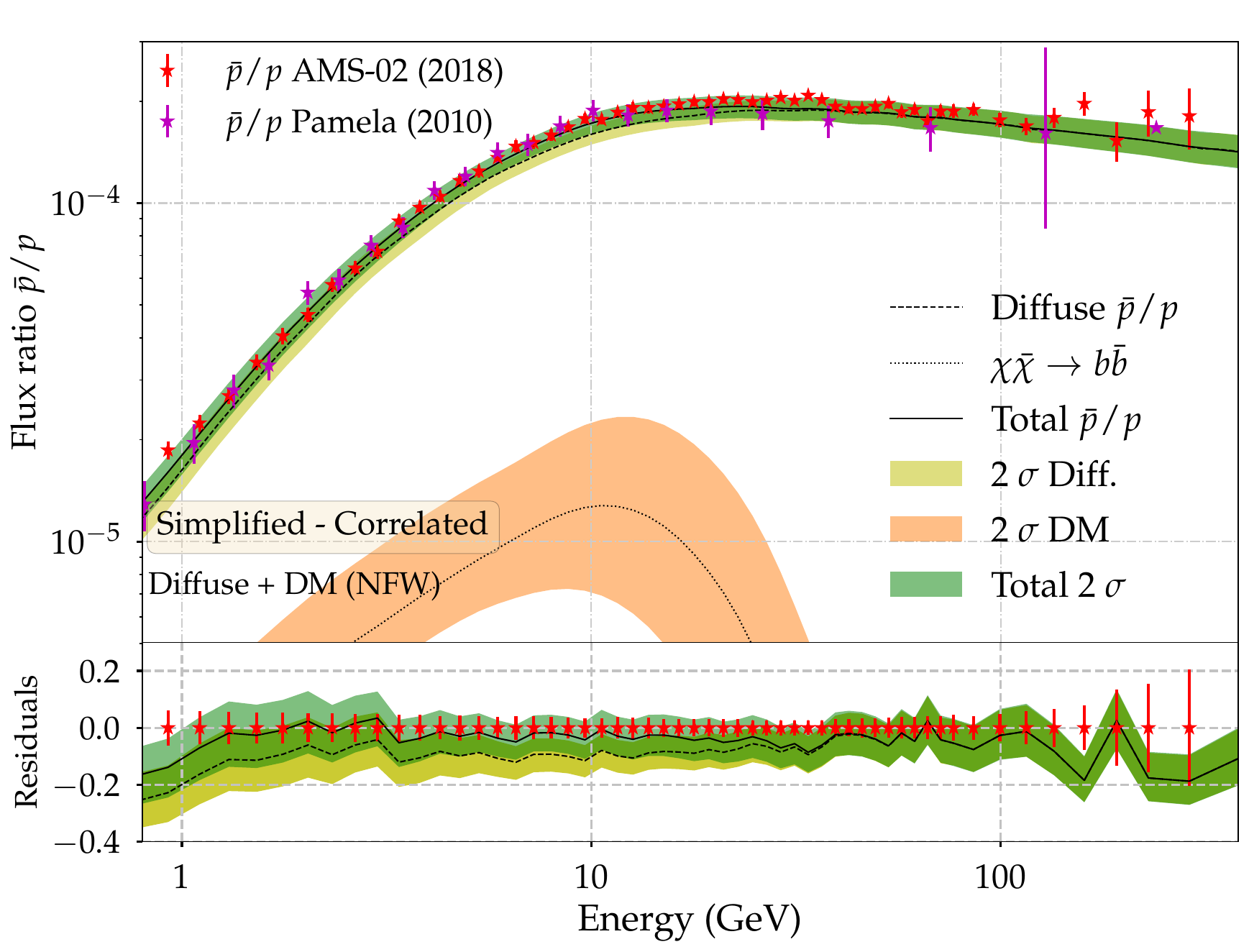} 
\hspace{0.07cm}
\includegraphics[width=0.485\textwidth]{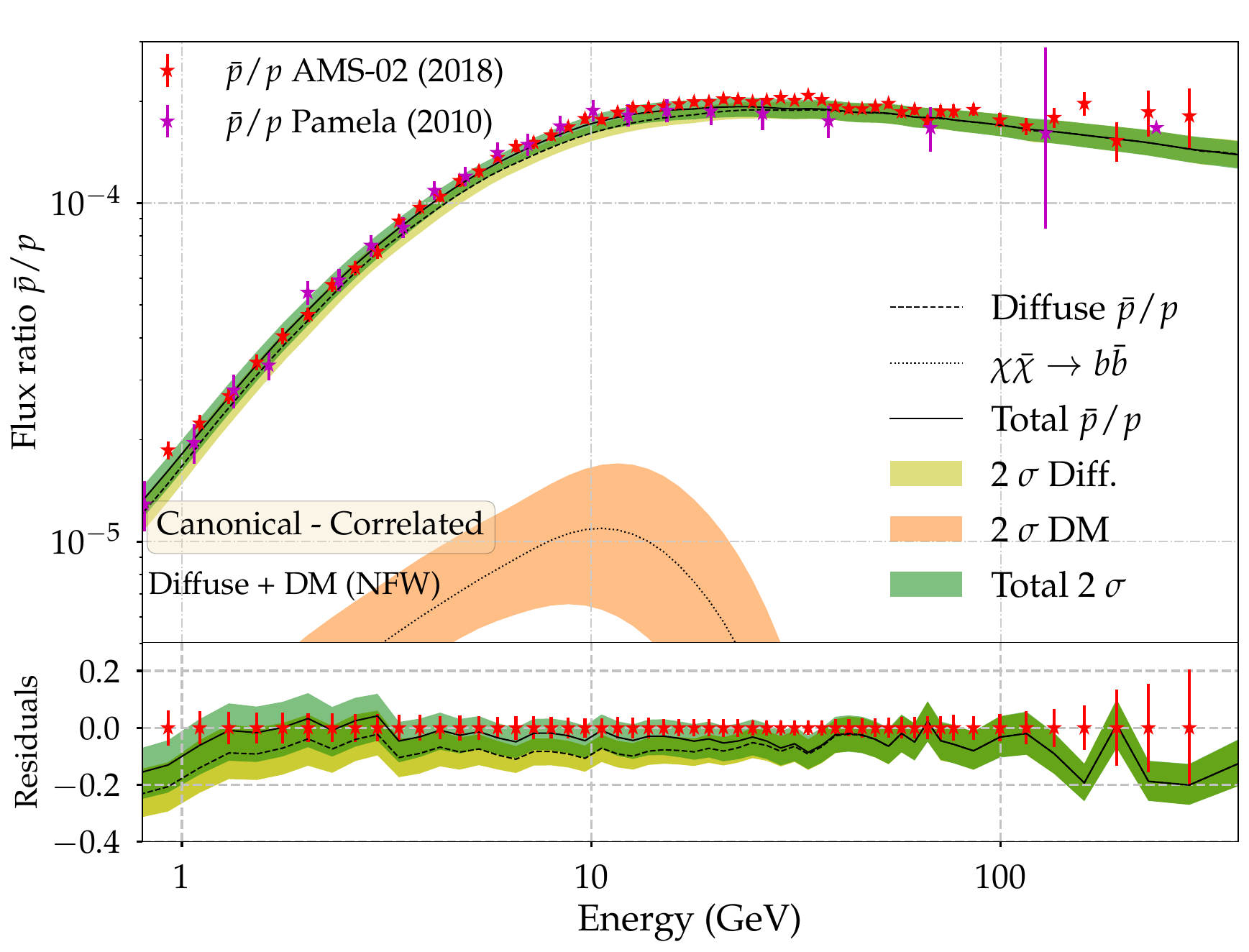}
\caption{$\bar{p}/p$ spectra evaluated in the scenario where the contribution from WIMP annihilation into $b\bar{b}$ final states is included, with the best-fit transport parameters obtained from the Canonical analysis assuming uncorrelated AMS-02 systematic errors (left panel) and accounting for the AMS-02 systematic error correlations (right panel). The statistical uncertainty in the determination of the propagation parameters (not including modulation uncertainties) is shown as a yellow band and the uncertainty related to the determination of the best-fit WIMP properties (mass and annihilation rate) is shown as an orange band. }
\label{fig:App_WIMP}
\end{figure}

In Figure~\ref{fig:App_WIMP} we show the best-fit $\bar{p}/p$ spectrum obtained in our Canonical analysis in the case where we do not include (left panel), or do include (right panel) correlations in the AMS-02 systematic errors. The best-fit DM  contribution to the $\bar{p}/p$ spectrum is lower in the case where we include error correlations. The best-fit mass and annihilation rate obtained in each analysis, as well as the local significance associated with the DM  contribution, are reported in Table~\ref{tab:DM_params}. 

Our results highlight the importance of including the correlations in the AMS-02 systematic errors when assessing the significance of a potential DM  signal. If we ignore the correlations, we find a relatively large local excess of $3.5-3.8\sigma$, compatible with early analyses~\cite{Cuoco, Cui2017,Cholis}. However, the local significance is reduced to $2.8-3\sigma$ once we account for the systematic error correlations in AMS-02 data. This general trend agrees with the findings of other recent studies~\cite{Reinert:2017aga, Heisig, WinklerDiMauro, Calore2021}. In our uncorrelated analyses, the best-fit mass of the DM  particle ranges from $51$~GeV to $100$~GeV when we include the covariance matrix for antiproton cross sections and when we only include a scaling factor, respectively. The best-fit annihilation rate is slightly below the thermal relic value for the Canonical analysis ($\langle \sigma v \rangle = 0.89\pm0.18 \times 10^{-26}$cm$^3$/s)  and compatible with the thermal relic value in the Simplified analysis ($\langle \sigma v \rangle = 2.01^{+0.4}_{-0.35} \times 10^{-26}$cm$^3$/s),  (see Table~\ref{tab:DM_params}).

We have, furthermore, computed the global significance of the excess by creating mock data samples to determine how often a local $2.8\sigma$ excess randomly occurs within the considered mass range (the look-elsewhere effect).\footnote{More details of the strategy are provided in Ref.~\cite{Reinert:2017aga}.} We obtain a global significance of only $1.8\sigma$, meaning that there is no statistically significant preference for a DM  contribution in agreement with~\cite{Reinert:2017aga,Heisig,WinklerDiMauro,Calore2021}. 
In the analyses that include AMS-02 error correlations, we get a best-fit mass of around $66$~GeV for an annihilation rate, $\langle \sigma v \rangle \sim 10^{-26}$~cm$^3$/s, somewhat below the thermal relic one. These values are quite similar to those found from the analysis of B/C and the $\bar{p}/p$ ratio (2015 AMS-02 dataset) using GALPROP in Ref.~\cite{Heisig}.

Other factors that can impact our results are the cross sections parameterizations that are employed, the evaluation of the correlation matrices for AMS-02 systematic errors, the low energy parameterization of the diffusion coefficient, the set of cosmic-ray observables included in the analysis, and the spatial dependence of the diffusion coefficient. For example, it was found by Ref.~\cite{Heisig} that including the antiproton flux instead of the $\bar{p}$/p ratio tends to decrease the significance of any antiproton excess. 

It is also worth noting that, in scenarios where we include the contribution from WIMP annihilation into $b\bar{b}$ final states, we still need to scale the cross sections of antiproton production by $\sim6-7\%$ (and as much as $14\%$ in the case that does not include the covariance matrix for antiprotons cross sections). However, the difference in the grammage needed to explain B, Be and Li compared to the one to explain antiprotons is reduced and the cross section scale factors for B, Be and Li are closer to unity (i.e. no scaling).
In general, we find that the best-fit cosmic-ray propagation parameters determined in models that include DM annihilation are not very different from those inferred in scenarios that include only $\bar{p}$ secondary production. This means that the determination of these parameters is mainly driven by the secondary ratios of B, Be and Li. A summary of the propagation parameters, scale factors, WIMP mass and $\left< \sigma v\right>$  found in these analyses is given in Figure~\ref{fig:Boxplot_DM}.
Future improvements in our understanding of the transport of CRs at low energy and solar modulation will allow us to pin down the astrophysical fluxes at low energy even more precisely and to further improve the sensitivity of our analysis to low-mass DM .

\subsubsection{Dark matter bounds}

\begin{figure}[!t]
\centering
\includegraphics[width=0.485\textwidth] {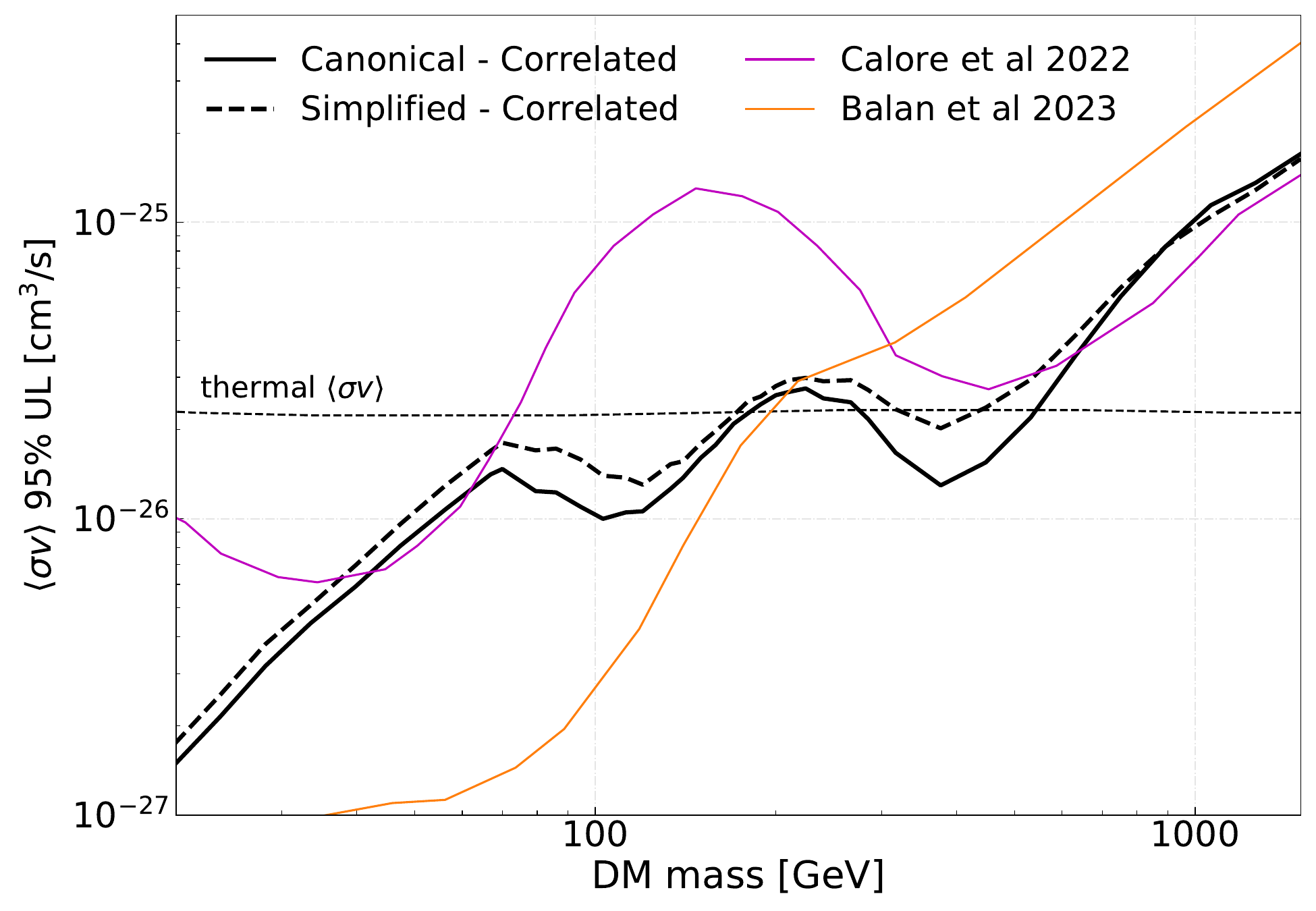} \hspace{0.2cm}
\includegraphics[width=0.485\textwidth] {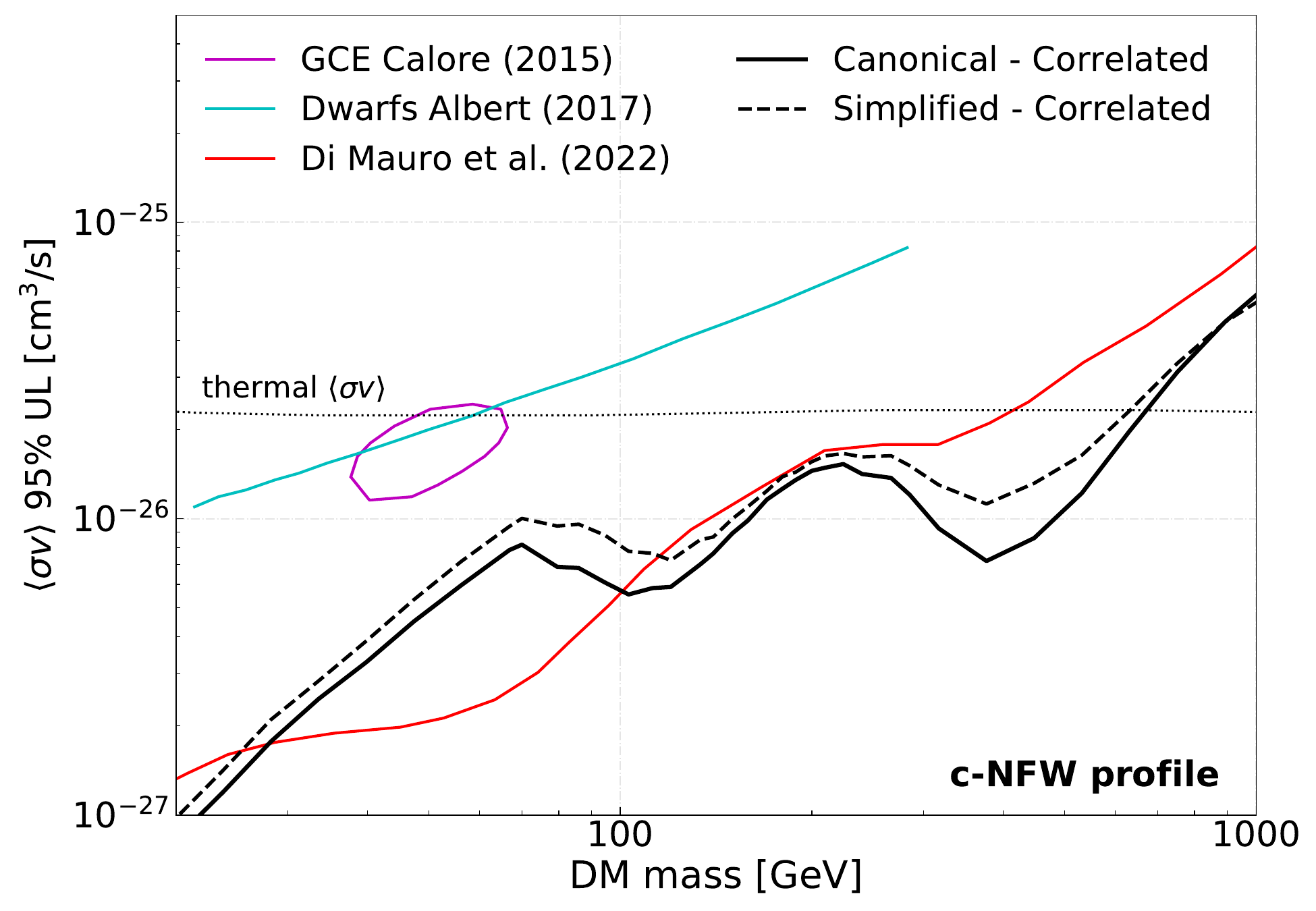} 
\caption{\textbf{Left panel}: 95\% confidence Upper limits on $\left< \sigma v\right>$ derived from our main analyses with a NFW profile, compared to those obtained by Refs.~\cite{Balan:2023lwg, Calore2021}. \textbf{Right panel}: Comparison of the upper limits obtained for a contracted NFW profile (see details in the text) with other limits derived for the same profile. In particular, we compare our limits with the $2\sigma$ contour representing the best-fit DM candidate reproducing the GCE in Ref.~\cite{Calore_GCE}, the constraints derived from dwarf spheroidal galaxies from Ref.~\cite{Albert_2017} and those from Ref.~\cite{WinklerDiMauro} for a contracted-NFW profile with a halo height of $H=4$~kpc (and the scale radius adjusted to $20$~kpc).}
\label{fig:DM_limits}
\end{figure}

Here, we report 95\% confidence upper limits on the dark matter annihilation cross-section, $\left< \sigma v\right>$, from our canonical analysis and compare our results to those of other groups. The limits are obtained using the same MCMC analysis described above, with the DM  cross-sections computed by logarithmically scanning the WIMP mass in $42$ mass bins, from $10$ to $1500$~GeV.
These bounds are shown in Fig.~\ref{fig:DM_limits}, where the left panel displays the limits obtained in our analysis including error correlations in the AMS-02 data, in comparison to the ones obtained by Refs.\cite{Calore2021, Kahlhoefer2021}.  
Over most of the considered mass range our limits fall between those of Ref.~\cite{Calore2021} and of Ref.~\cite{Balan:2023lwg}. The differences compared to those previous works arise due to the advancements of our CR analysis, described in Section~\ref{sec:Comparison}.
We finally remark that these limits allow us to rule out the thermal relic cross sections for masses below $\sim200$~GeV in the $b\bar{b}$-channel.

In the right panel of Figure~\ref{fig:DM_limits}, we show the bounds we obtain for dark matter profiles that follow a contracted NFW profile (favored by the DM  interpretation of the galactic center gamma ray excess) with $\gamma=1.2$ and a scale radius of $20$~kpc, compared to those obtained from analyses of the antiproton cross sections in Ref.\cite{WinklerDiMauro}. The $2\sigma$ contour represents the best-fit DM candidate reproducing the GCE derived by Ref.~\cite{Calore_GCE} and the constraints derived from dwarf spheroidal galaxies in Ref.~\cite{Albert_2017}. In this plot, we show our upper limits on the annihilation rate for our Canonical and Simplified analyses, including correlations in the AMS-02 errors. 

As we can see from this comparison, the DM  interpretation of the GCE is in tension with respect to these antiproton analyses for the $\bar{b}b$ channel (even in the uncrorrelated case) as was shown also in Ref.~\cite{WinklerDiMauro}. In fact, the best-fit candidate found by Ref.~\cite{Calore_GCE} (which is one of the analyses leading to a lower annihilation rate for the DM candidate fitting the GCE) is around a factor of $3$ above our $95\%$ confidence limits. However, we remark that the exploration of other annihilation channels and configurations of the DM profile (mainly contraction index and scale radius), which is rather uncertain close to the Galactic center, is needed to clearly state that the DM explanation for the GCE and the AMS-02 antiproton data are inconsistent (for example, Ref.~\cite{WinklerDiMauro} found that both are compatible in the $\mu^{\pm}$ channel). Moreover, uncertainties relating to inhomogeneous cosmic-ray propagation, {\it e.g.,} a strong convective wind near the galactic center, could change the relative intensity of the $\gamma$-ray and antiproton signals.
Then, we show the limits from Ref.~\cite{Albert_2017}, which were shown to be robust to different DM profiles. This also allows us to illustrate the fact that, when using a contracted NFW, antiproton measurements can certainly be the leading observable to constrain the existence of DM. 
The bounds obtained in the uncorrelated cases, as well as other analyses, are discussed and shown in App.~\ref{sec:Alternative}. In particular, we show the impact of considering uncorrelated AMS-02 systematic uncertainties in the DM bounds by comparing with those including correlations in Fig.~\ref{fig:Alternative_limits} (left panel).

In view of these results, we conclude that,
even with the simple propagation setup generally assumed, the spectra of the different secondary CRs agree with pure secondary production, and do not require any additional antiproton production from WIMPs. These analyses allow us to limit the contribution from a WIMP particle of mass below $\sim1$~TeV to be, at most, a $\sim15\%$ of the secondary contribution, for the $\bar{b}b$ channel. 
This highlights the great need to further improve cross sections through accelerator measurements in order to (along with the development of a more robust astrophysical modeling) improve our searches for small signals on top of large backgrounds in the antiproton spectrum. For instance, our rescaling factors on nuclear and antiproton cross sections should be revised with future accelerator data.

Until the uncertainties in the relevant CR cross sections and in the CR propagation are significantly reduced it will be difficult to unambiguously identify a DM signal with CR antiprotons -- even with more precise AMS-02 data. Therefore, it is of great importance to investigate DM signals in complementary CR channels with lower astrophysical backgrounds.
In a companion paper, we make use of our antiproton results and update the predictions for antinuclei (antideuteron and antihelium) detection, using newly derived cross sections for their production from CR interactions and WIMP annihilation and discussing the implications of the updated predictions in view of the preliminary detected events reported by the AMS-02 collaboration.

\begin{table}[t!]
\centering
\resizebox*{\columnwidth}{0.10\textheight}{
\begin{tabular}{|c|c|c|c|c|c|c|c|}
\hline & \multicolumn{3}{|c|}{Correlated AMS-02 errors} &  \multicolumn{3}{|c|}{Uncorrelated AMS-02 errors} \\  

\hline & M$_{WIMP}$ (GeV) & $\left< \sigma v \right>$ ($10^{-26}$ cm$^3$/s) & local $\sigma$ & M$_{WIMP}$ (GeV) & $\left< \sigma v \right>$ ($10^{-26}$ cm$^3$/s) & local $\sigma$ \\ 
\hline
\centering \textbf{Canonical} & \textbf{66.28$^{+3.18}_{-4.9}$}  & \textbf{0.99$^{+0.2}_{-0.21}$} &  \textbf{2.8} &  51.1$^{+3.14}_{-3.89}$  & 0.89$^{+0.18}_{-0.18}$  & 3.5 \\ \hline 

\centering Simplified & 66.95$^{+3.43}_{-5.37}$  & 1.18$^{+0.26}_{-0.27}$ & 3.03 &  100.64$^{+9.4}_{-9.05}$  & 2.01$^{+0.40}_{-0.35}$  & 3.8 \\ \hline 
\end{tabular}
}
\caption{Best-fit parameters characterizing the WIMP properties, namely mass and annihilation rate, along with the local significance found for the WIMP contribution. In particular, the median of the PDF and the $1\sigma$ error are shown. The reported error corresponds to the $1\sigma$ uncertainty obtained from the PDF fixing the other parameters to their best-fit values.}
\label{tab:DM_params}
\end{table}

\section{Discussion and conclusions}
\label{sec:conc}

Over the last decade, many studies have investigated the spectra of CR antiparticles in order to search for hints of new physics. Although there have been indications of significant anomalies in the $\bar{p}$ spectrum that indicate a possible signal from DM, no clear evidence has yet been uncovered. In this work, we analyzed the spectra of the light secondary CR species B, Be, Li in combination with $\bar{p}$ in a scenario where antiprotons can also be produced by annihilation of a WIMP into $b\bar{b}$ final states, completing and extending our previous work. This constitutes the current most complete derivation of DM bounds from AMS-02 antiproton data, characterized by an analysis that combines antiprotons with the rest of light secondary CRs, uses the updated DRAGON2 cross sections for spallation interactions, and accounts for uncertainties in the cross sections parametrizations and possible correlations in AMS-02 data, to search for WIMP signals in the antiproton spectrum.  

We find that the antiproton spectrum is compatible with pure astrophysical production and obtain a local significance of $\lesssim 3\sigma$ for any additional antiproton component from DM  annihilation. This result is obtained within our main analysis that includes correlations between the AMS-02 systematic errors. We have tested that this corresponds to a global significance of $<2\sigma$ when accounting for the trials factor from the scan of DM  masses, indicating that there is no strong statistical preference for a DM  contribution. 

Regarding the differences obtained considering and neglecting correlations in the AMS-02 errors, we observe that the significance in our main analyses vary by around $0.7 \sigma$, being of $\sim3.5-3.8\sigma$ (local) in the uncorrelated case. This shows the importance of including correlations in these analyses in order to prevent from overestimating a potential DM signal.



We have reported DM  bounds and showed the effect of treating cross sections uncertainties in different ways, finding that our bounds are compatible with other recent analyses of the 2018 antiproton data-set. In particular, our analyses allow us to consistently rule out the thermal relic cross sections for WIMP masses below $\sim200$~GeV. 
The mass of the DM  particle that produces the largest improvement in the antiproton spectrum is $66$~GeV, which agrees well with previous works, and is also interesting because it closely corresponds to the best-fit mass for DM  explanations of the galactic center $\gamma$-ray excess. However, the bounds on the annihilation rate obtained in our analysis are in tension with the best-fit cross-sections for DM explanations of the GCE in the $\bar{b}b$ channel. 

Let us remind the reader that our analysis takes into account cross section uncertainties (on the secondary production of antiprotons and nuclei) through overall scaling factors and also includes energy-dependent uncertainties on antiproton cross sections through covariance matrices. Given the complexity of our fit, which involves a much larger set of CR species compared to previous analyses, incorporating the energy scaling of cross section uncertainties for each isotope of B, Be and Li goes beyond the scope of this work. We note, however, that the significance of the excess is expected to further (slightly) reduce once the energy-scaling of cross section uncertainties for these secondary CRs is taken into account (see discussion in Refs.~\cite{Reinert:2017aga,Heisig,WinklerDiMauro,Calore2021}). A more detailed investigation of this aspect is left for future work. 

Finally, we note that it would be interesting to generalise our analyses to account for possible spatial dependence of the diffusion coefficient, something that could affect slightly both, the spectra of particles produced by CR interactions and the DM  signal at Earth. This study is left for a next work. Additionally, in a companion paper, we will use our constraints to re-evaluate the expected fluxes of antideuteron and antihelium at Earth. The version of the code employed in this and our companion work is publicly available at \url{github.com/tospines/Customised-DRAGON-versions/tree/main/Custom_DRAGON2_v2-Antinuclei}.

\acknowledgments

PD and TL are supported in part by the European Research Council under grant 742104 and the Swedish National Space Agency under contract 117/19. TL is also supported by the Swedish Research Council under contracts 2019-05135 and 2022-04283. MW acknowledges support by the U.S.\ Department of Energy, Office of Science, Office of High Energy Physics program under Award Number DE-SC-0022021 and by the Swedish Research
Council (Contract No.\ 638-2013-8993).
This project used computing resources from the Swedish National Infrastructure for Computing (NAISS) under project Nos. 2021/3-42, 2021/6-326, 2021-1-24 and 2022/3-27 partially funded by the Swedish Research Council through grant no. 2018-05973.

\bibliographystyle{apsrev4-1}
\bibliography{biblio}

\appendix

\section{DRAGON2 cross sections for spallation interactions of B, Be and Li}
\label{sec:appendixA}

In this appendix, we show different models of direct cross sections for spallation reactions used in the literature compared to experimental data in some of the most important reaction channels for the production of the light secondary CRs B, Be and Li. Experimental data are taken from various experiments and authors (see Refs.~\cite{DRAGON2-2, Evoli:2019wwu}): EXFOR (Experimental Nuclear Reaction Data)\footnote{\url{https://www-nds.iaea.org/exfor/exfor.dhtm}}, the {\tt GALPROP} cross-section database ($isotope\_cs.dat$) and from various publications and experiments (Bodemann1993, Davids1970, Fontes1977, Korejwo1999, Korejwo2002, Moyle1979, Olson1983, Radin1979, Read1984, Roche1976, W90, W98a and Zeitlin2011). It is important to remind that the FLUKA cross sections are computed by means of the FLUKA code~\cite{delaTorreLuque:2022vhm} while the other models constitute different parameterizations fitted to data existent at the time of release~\cite{webber2003updated, silberberg1998updated, moskalenko2005propagation}. 

Full details about the different parameterizations can be found in Refs.~\cite{delaTorreLuque:2022vhm, Luque:2021joz, GALPROPXS, GenoliniRanking} and references therein.
\begin{figure*}[!hb]
\label{fig:XSB}
\centering
\includegraphics[width=0.35\textwidth,clip] {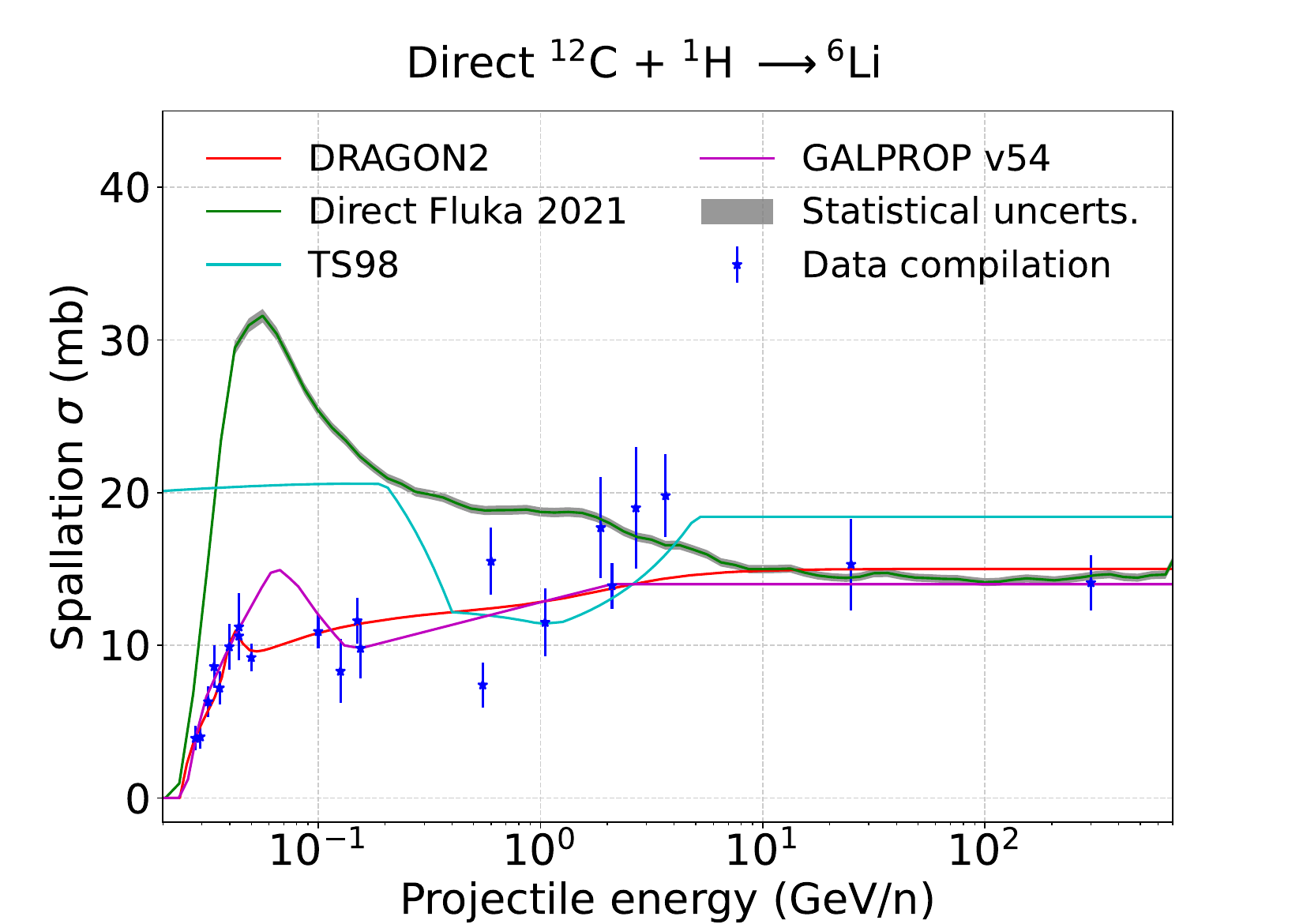} \hspace{-0.65cm}
\includegraphics[width=0.35\textwidth,clip] {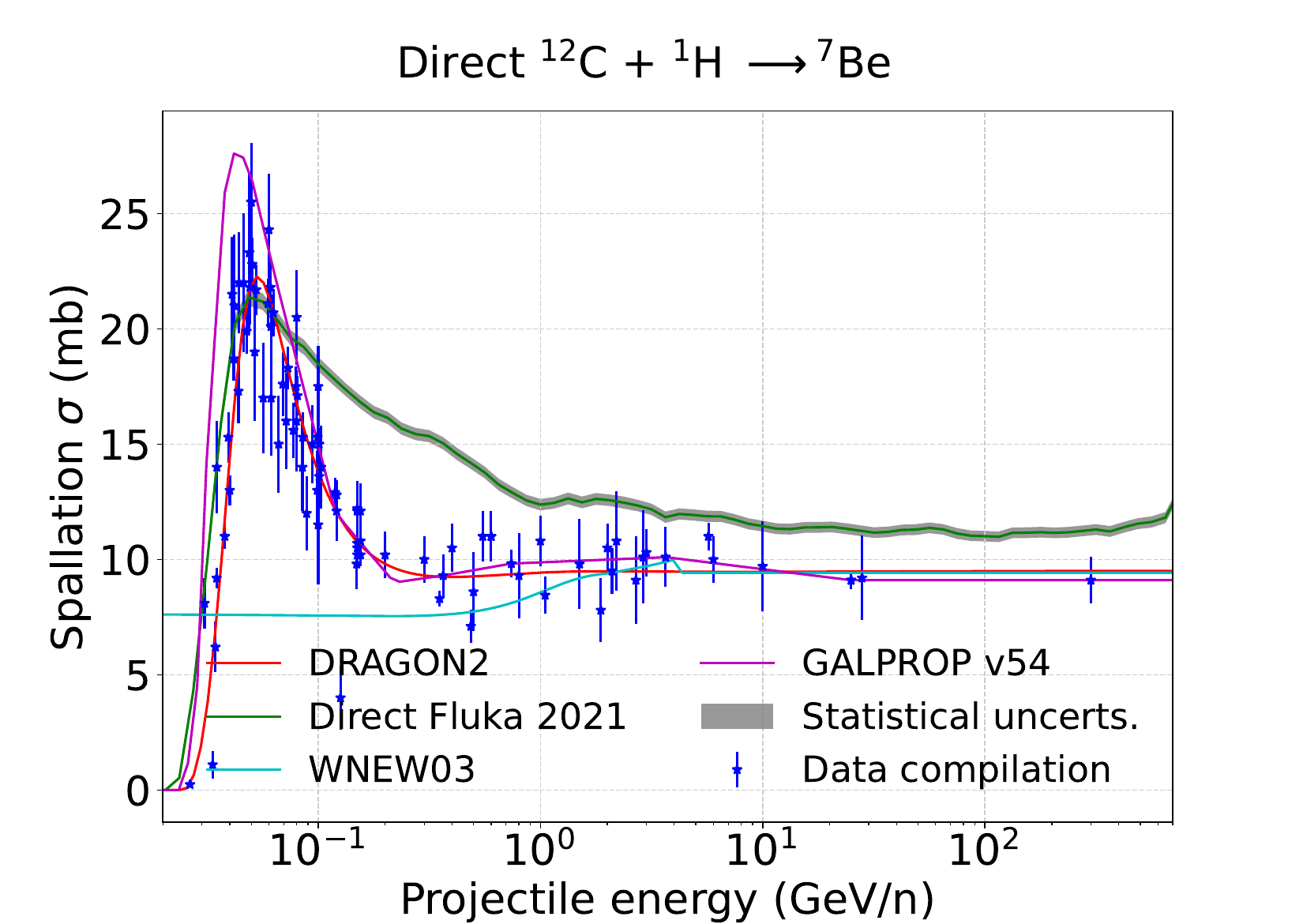} \hspace{-0.65cm}
\includegraphics[width=0.35\textwidth,clip] {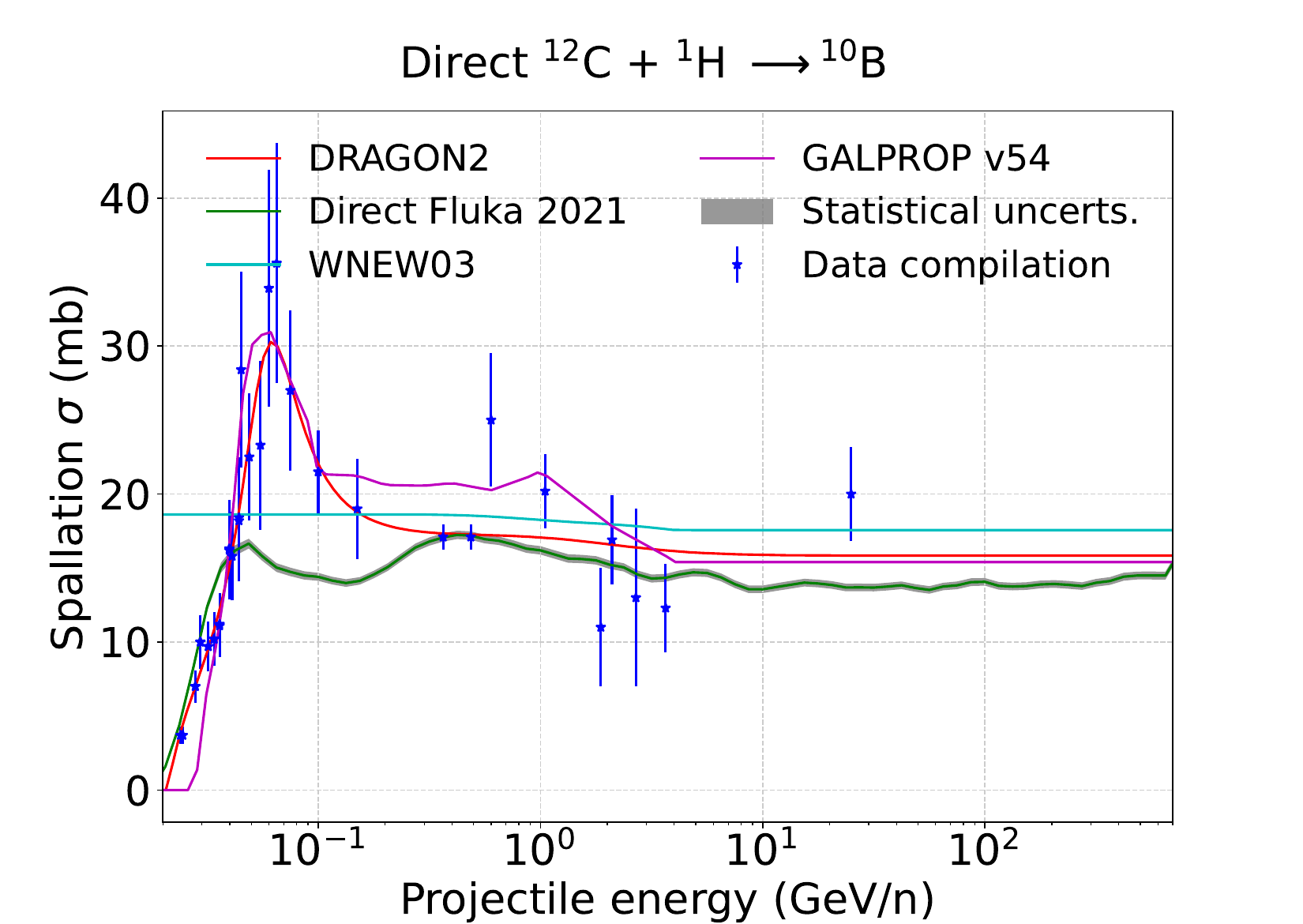} 

\includegraphics[width=0.35\textwidth, clip] {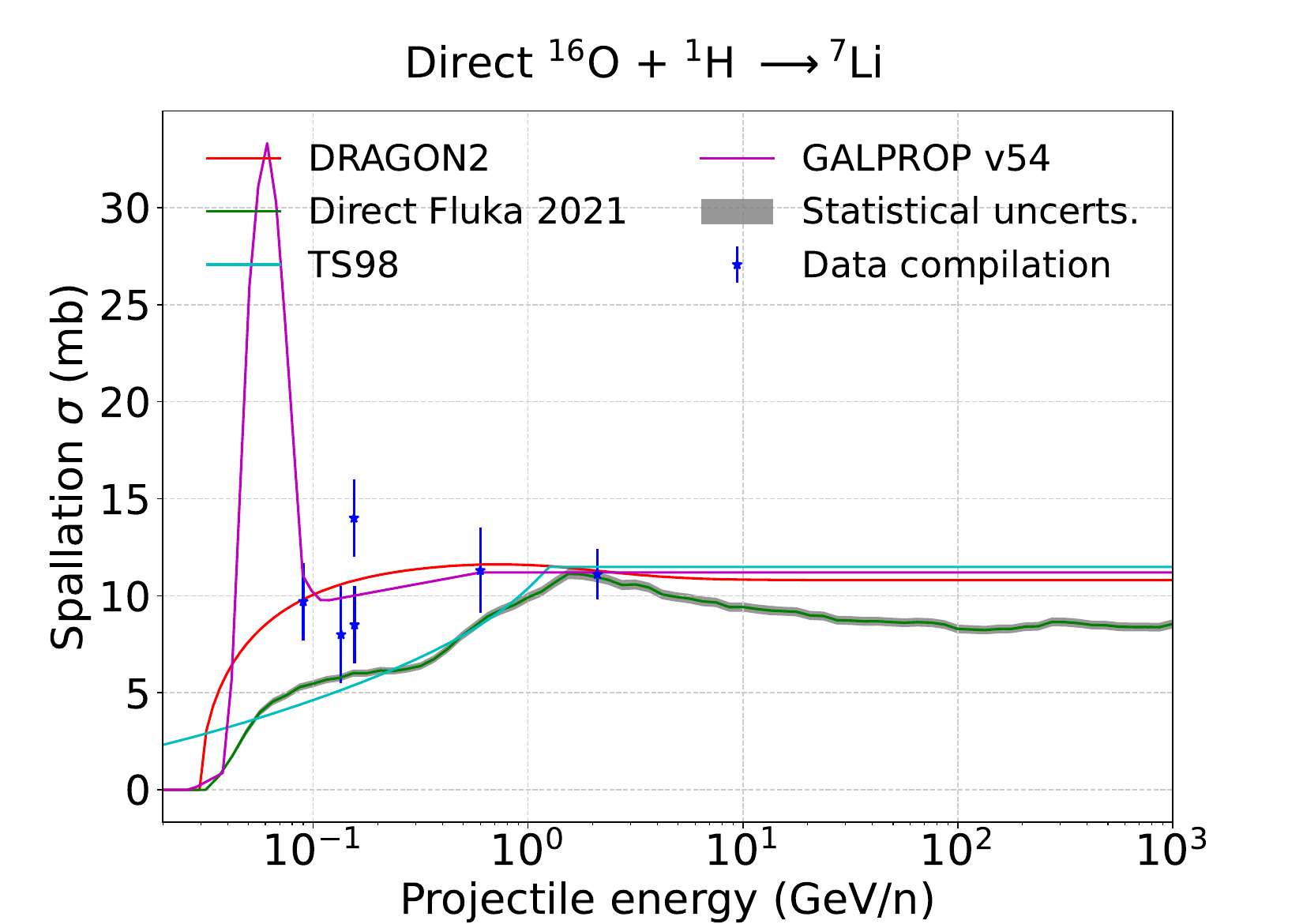} \hspace{-0.65cm}
\includegraphics[width=0.35\textwidth, clip] {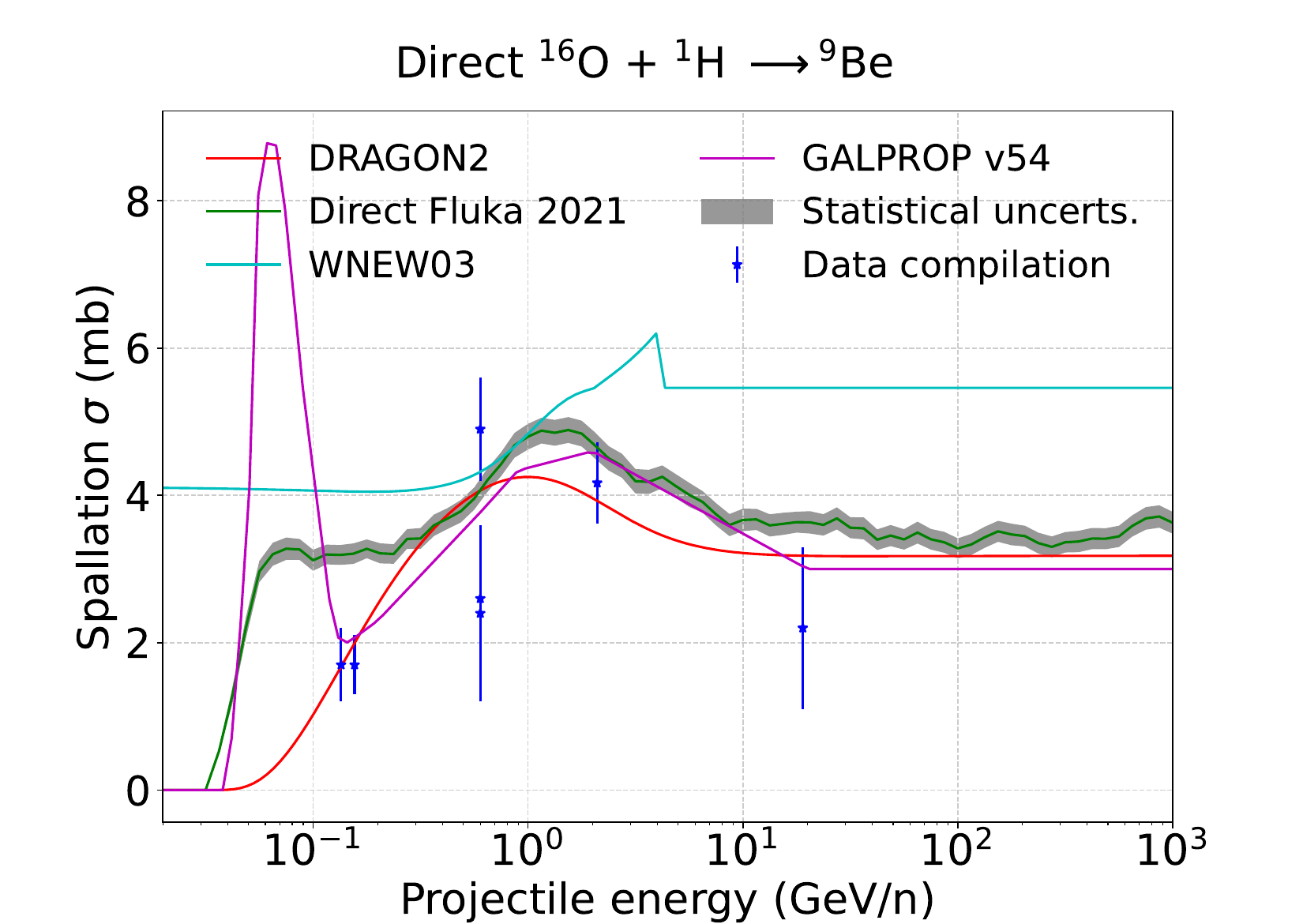} \hspace{-0.65cm}
\includegraphics[width=0.35\textwidth, clip] {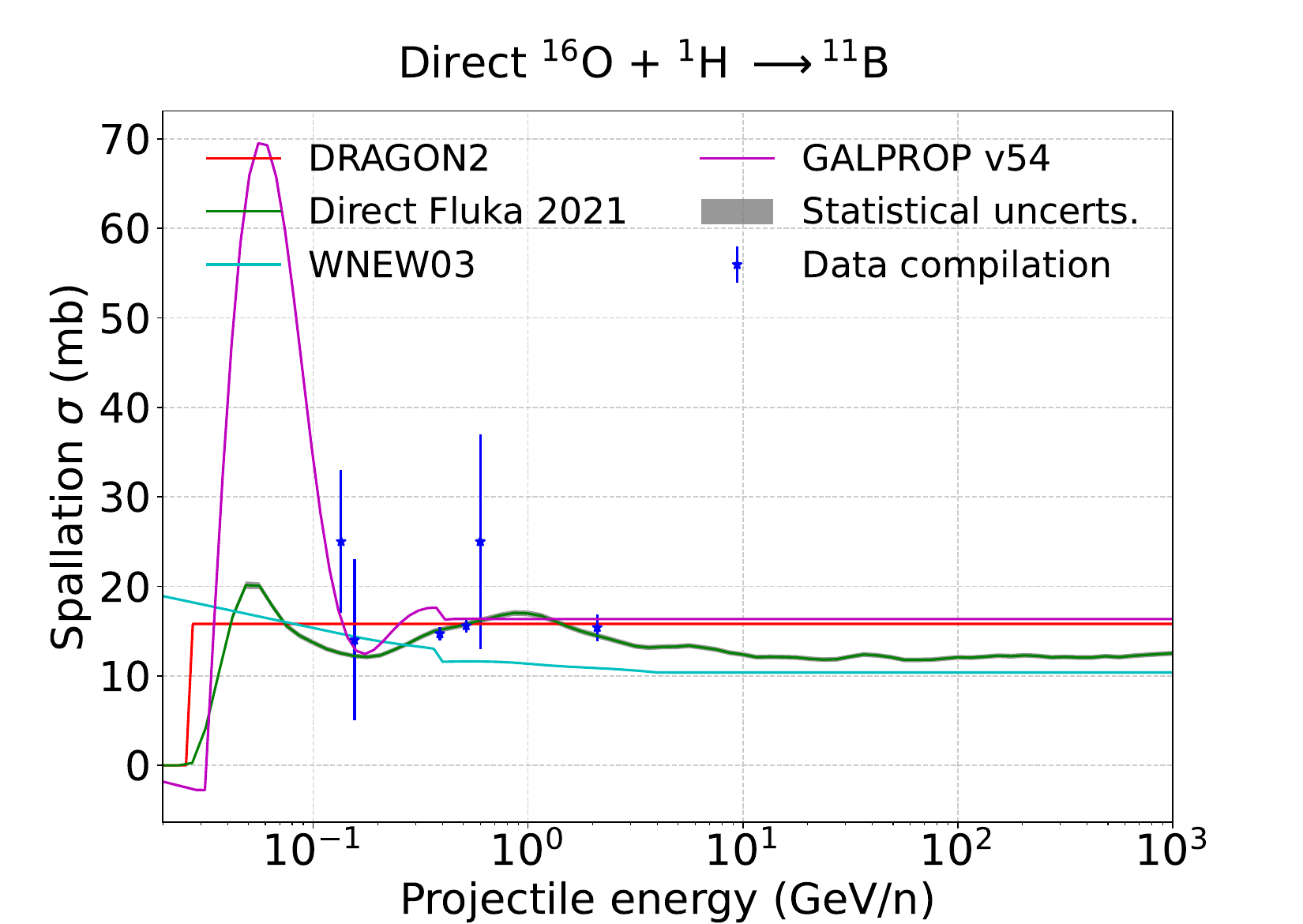} 
\caption{Most used cross sections models in literature compared to available experimental data for the production of different isotopes of B, Be and Li coming from $^{12}$C (top row) and $^{16}$O (bottom). The band of statistical uncertainties is related to the determination of the FLUKA cross sections (see Ref.~\cite{delaTorreLuque:2022vhm} for more details).}
\end{figure*} 

\section{Other observables included in the analyses}
\label{sec:appendixB}

\begin{figure}[htb!]
\centering
\includegraphics[width=0.48\textwidth] {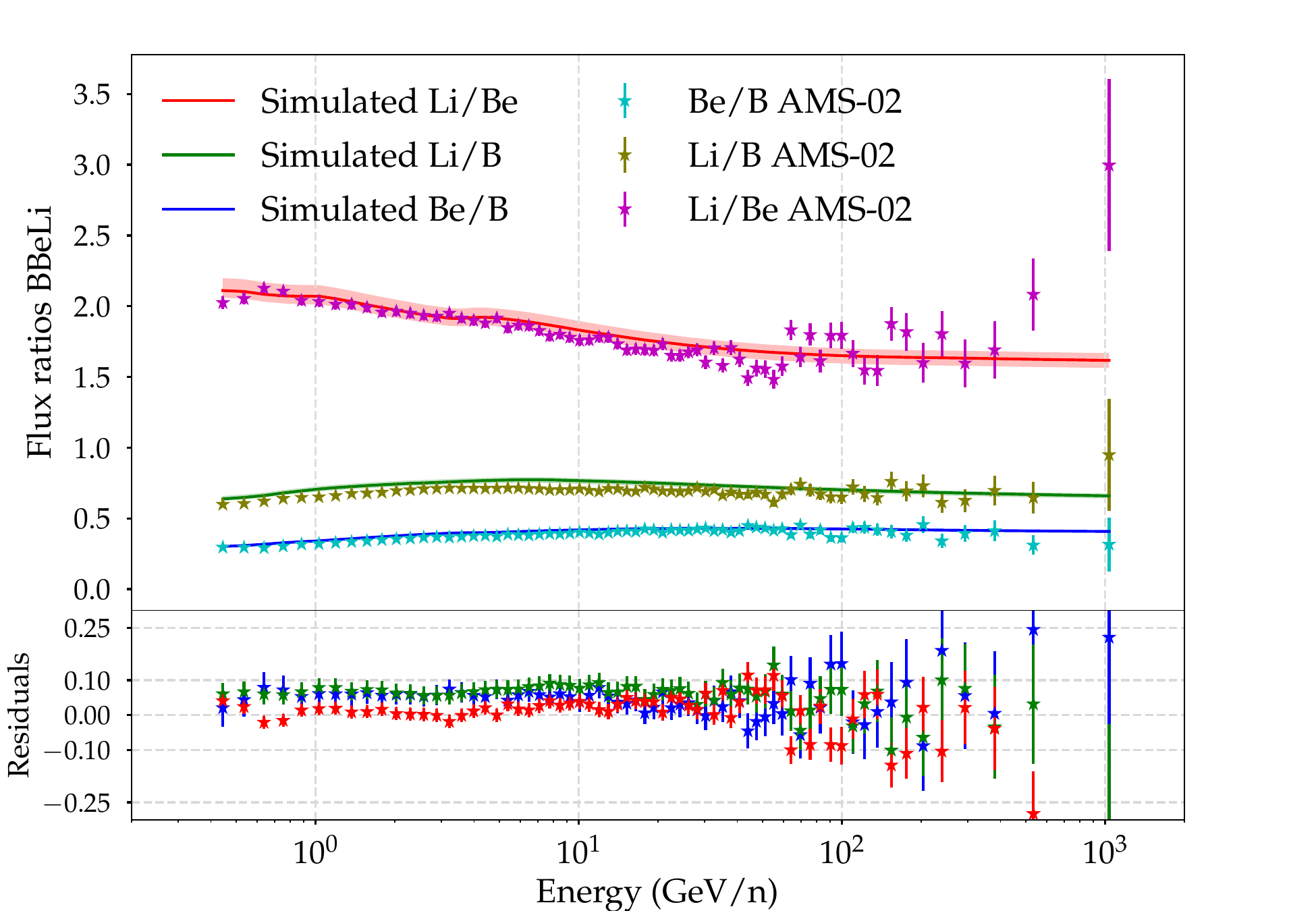}
\includegraphics[width=0.48\textwidth,clip] {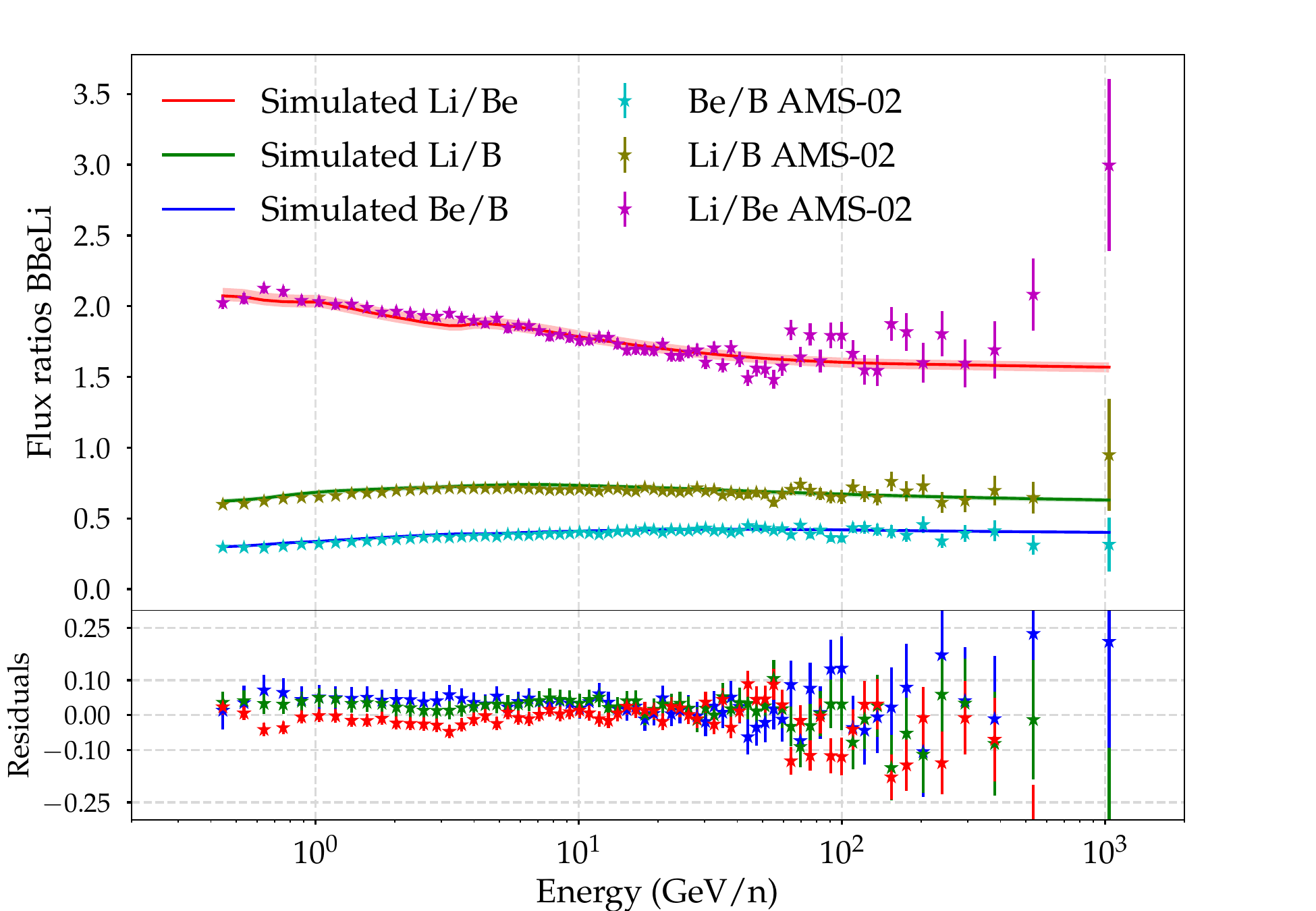} 
\vspace{-0.1cm}
\includegraphics[width=0.48\textwidth,height=0.215\textheight,clip] {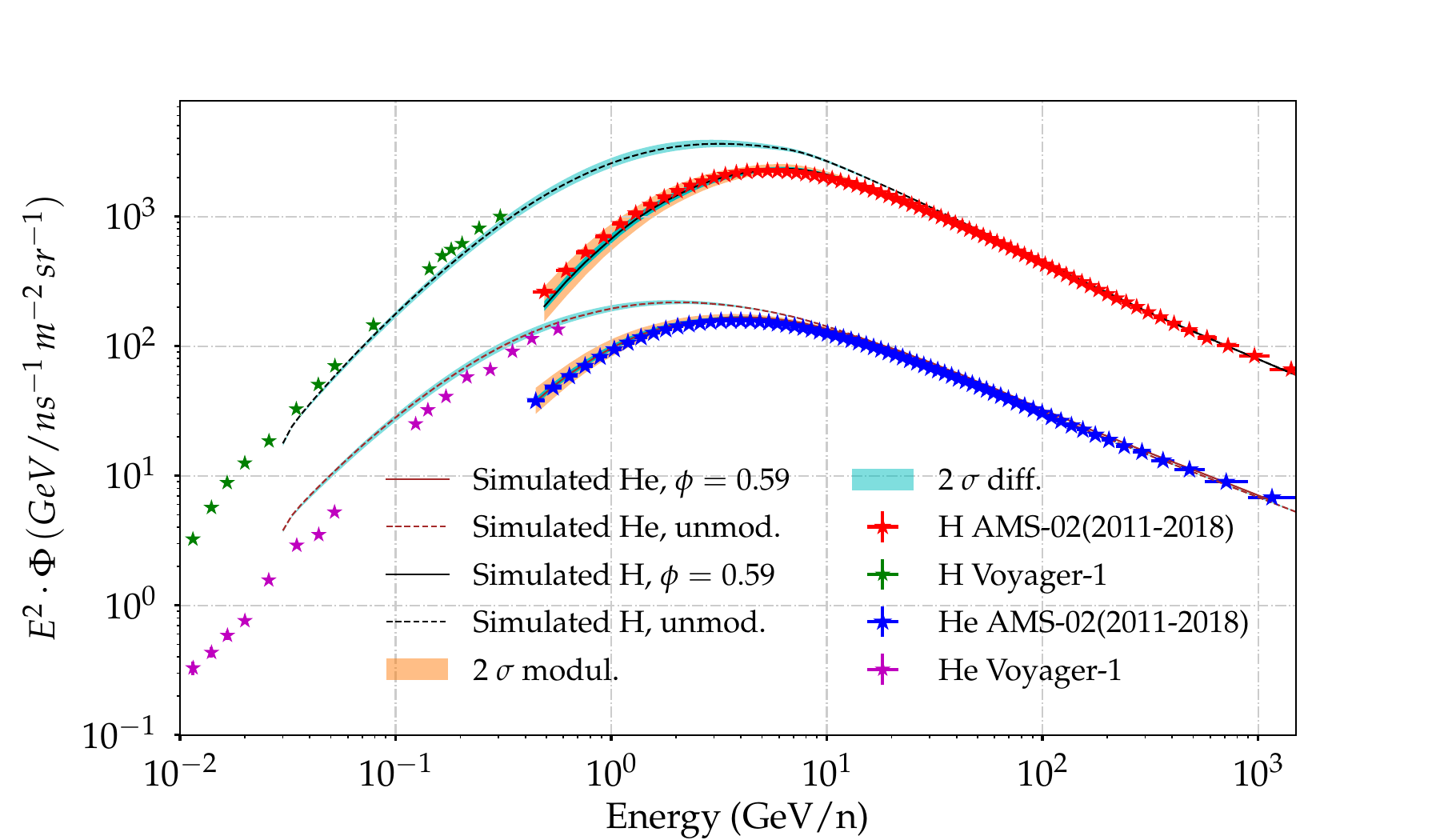} 
\includegraphics[width=0.48\textwidth, height=0.22\textheight,clip] {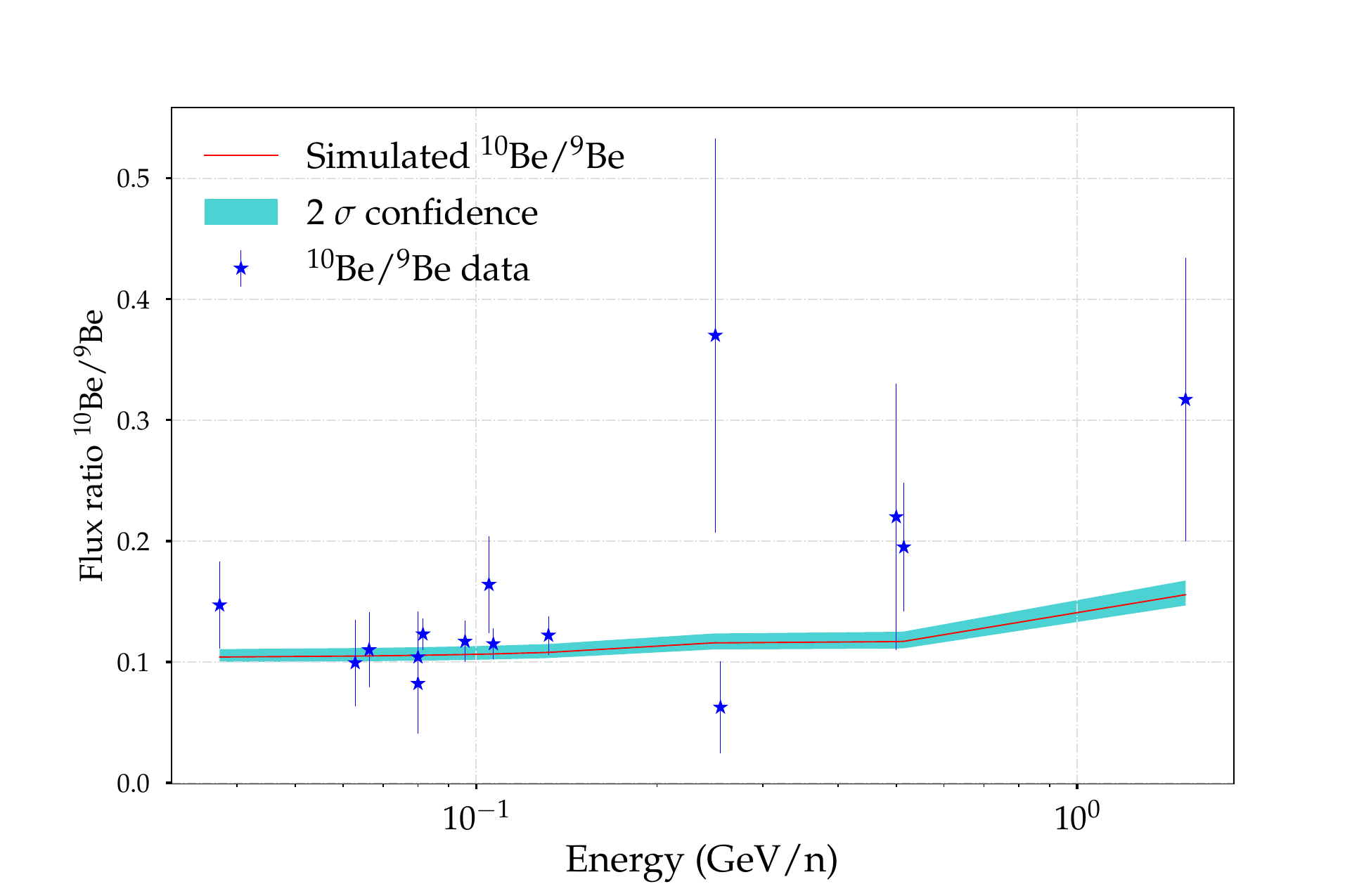}
\caption{Top row: Comparison of the different secondary-to-secondary rations of B, Be and Li evaluated from the parameters obtained assuming correlations in AMS-02 errors (top-left) and without correlations (top-right), compared to the AMS-02 data. The $2\sigma$ uncertainty from the determination of the diffusion parameters is also reported as a band for each ratio not including modulation uncertainties
Bottom-left: Modulated (solid lines) and unmodulated (dashed lines) fluxes of H and He compared to AMS-02 and Voyager-1 data. The $2\sigma$ uncertainty bands from the determination of the propagation parameters and the Fisk potential are also shown. 
Bottom-right: Flux ratios of Li/Be, Li/B and Be/B compared to AMS-02 data, including $2\sigma$ uncertainty bands from the determination of the propagation parameters. Bottom-right panel: $^{10}$Be/$^9$Be flux ratio with the uncertainties related to propagation parameters determination, compared with data from the ACE, IMP, ISEE, ISOMAX, Ulysses and Voyager experiments.}
\label{fig:Others}
\end{figure}

In this appendix we report the predicted fluxes and ratios evaluated from the best-fit parameters obtained in our canonical analysis. In the top panels of Fig.~\ref{fig:Others} we show a comparison of the different secondary-to-secondary rations of B, Be and Li evaluated from the parameters obtained assuming correlations in AMS-02 errors (top-left panel) and without correlations (top-right panel), compared to the AMS-02 data. The $2\sigma$ uncertainty from the determination of the diffusion parameters is also reported as a band for each ratio not including modulation uncertainties. As we see, in both kind of analyses, the these models achieve a good reproduction of the experimental data, even at the $1\sigma$ level.

In the bottom-left panel of this figure, we show the fluxes of H and He compared to AMS-02 (for the modulated fluxes) and Voyager-1 data (for the unmodulated fluxes) as well as the $2\sigma$ uncertainty bands from the determination of the propagation parameters and the Fisk potential ($\Delta\phi_{2\sigma} \sim 0.12$). The obtained flux of these primary CRs is roughly the same in every analysis since the injection parameters for primary CRs are left free in every analysis.
Finally, the bottom-right panel shows the $^{10}$Be/$^9$Be flux ratio with the uncertainties related to the determination propagation parameters. In this panel, we have compared our results with data from the ACE~\cite{ACEBe}, IMP~\cite{IMP1, IMP2}, ISEE~\cite{ISEE}, ISOMAX~\cite{Hams_2004}, Ulysses~\cite{UlysesBe} and Voyager~\cite{VoyagerMO} experiments. 

\section{Comparison of the bounds derived in this work}
\label{sec:Alternative}

\begin{figure}[!h]
\centering
\includegraphics[width=0.48\textwidth] {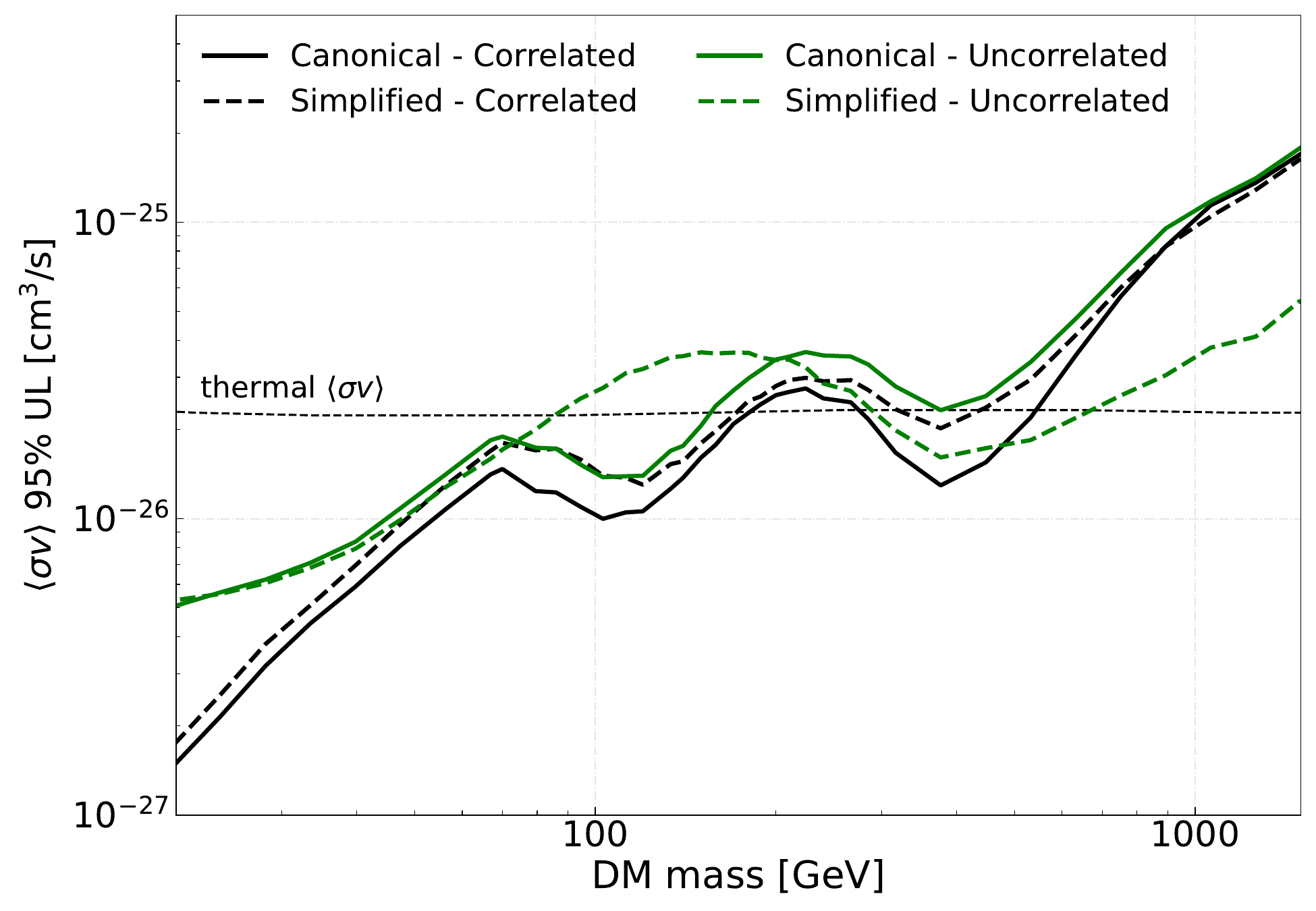} 
\includegraphics[width=0.48\textwidth] {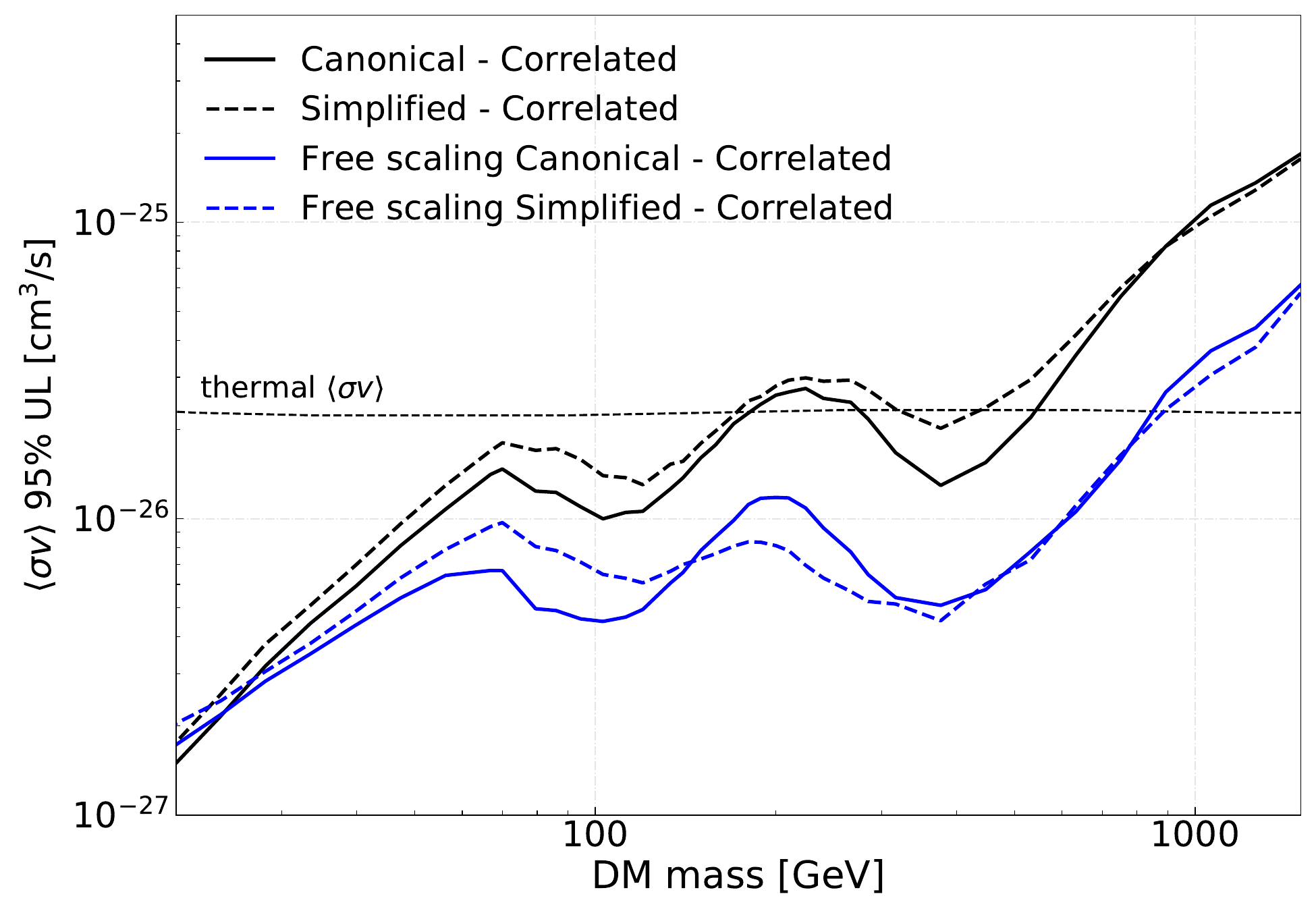} 
\caption{Comparison of the 95\% confidence upper limits on $\left< \sigma v\right>$ derived from the different analyses performed in this work. In the left panel, we compare our bounds with those obtained by Refs.~\cite{Balan:2023lwg, Calore2021}, for a NFW profile. In the right panel, we show the bounds derived from our alternative analyses, where we left free the scaling factor for B, Be, Li and $\bar{p}$ cross sections and correlations in AMS-02 errors are considered. These are compared with the results from our Canonical and Simplified analyses.
In both panels, we show the limits from the analyses including correlations in the AMS-02 systematic errors. The bounds from the Canonical analysis are represented by solid lines while those from the Simplified analyses are represented by dashed lines.}
\label{fig:Alternative_limits}
\end{figure}

In this appendix, we show, in Fig.~\ref{fig:Alternative_limits}, the bounds derived in the analyses assuming no correlations in AMS-02 errors (in the left panel) and other bounds obtained allowing the cross-section scale factors to float freely. In the left panel of this figure, the limits obtained from the uncorrelated analyses are shown as green lines (solid when including the cross sections covariance matrix for antiprotons and dashed one for the Simplified analysis) and those obtained including AMS-02 error correlations as black lines. 

Additionally, we performed an alternative analysis where we left the scaling factors to be free to scale (i.e. no penalty factor associated to the cross sections scale factors of B, Be, Li and $\bar{p}$). We applied this to the case where we include the covariance matrix for $\bar{p}$ cross sections uncertainties and when we only use a scale factor, always considering error correlations in AMS-02 data. For the former, we find a best-fit mass of around $50$~GeV, with a best-fit $\langle\sigma v \rangle= 3\times10^{-27}$~cm$^3$/s and an antiproton cross section scaling of $\sim4\%$. Meanwhile, in the analysis that does not include the energy dependent uncertainties in the antiproton cross sections, the best-fit mass is around $70-80$~GeV with a best-fit $\langle\sigma v \rangle= 1.1\times10^{-26}$~cm$^3$/s and an antiproton cross section scaling of $\sim11\%$.

In both cases, the significance of any antiproton excess is below $1\sigma$. This means that leaving more freedom to our analysis, we find a set of reasonable propagation parameters and scale factors that lead to simultaneous fits of the B, Be, Li and $\bar{p}$ spectra with a negligible preference for any additional DM contribution.
The derived bounds at $95\%$ C.L. from the analysis with free cross section scale are shown in the right panel of Fig.~\ref{fig:Alternative_limits}, as blue lines, and compared to those derived in our main analyses. These analyses show that leaving free the scaling of the cross sections of the secondary CRs B, Be, Li and $\bar{p}$, we get best-fit propagation values with even lower significance favoring a DM contribution ($<1\sigma$), and much more stringent limits.

\section{Summary of the MCMC results: propagation parameters}
\label{sec:diff_params}
\begin{figure}[h]
\centering
\includegraphics[width=0.47\textwidth,clip]{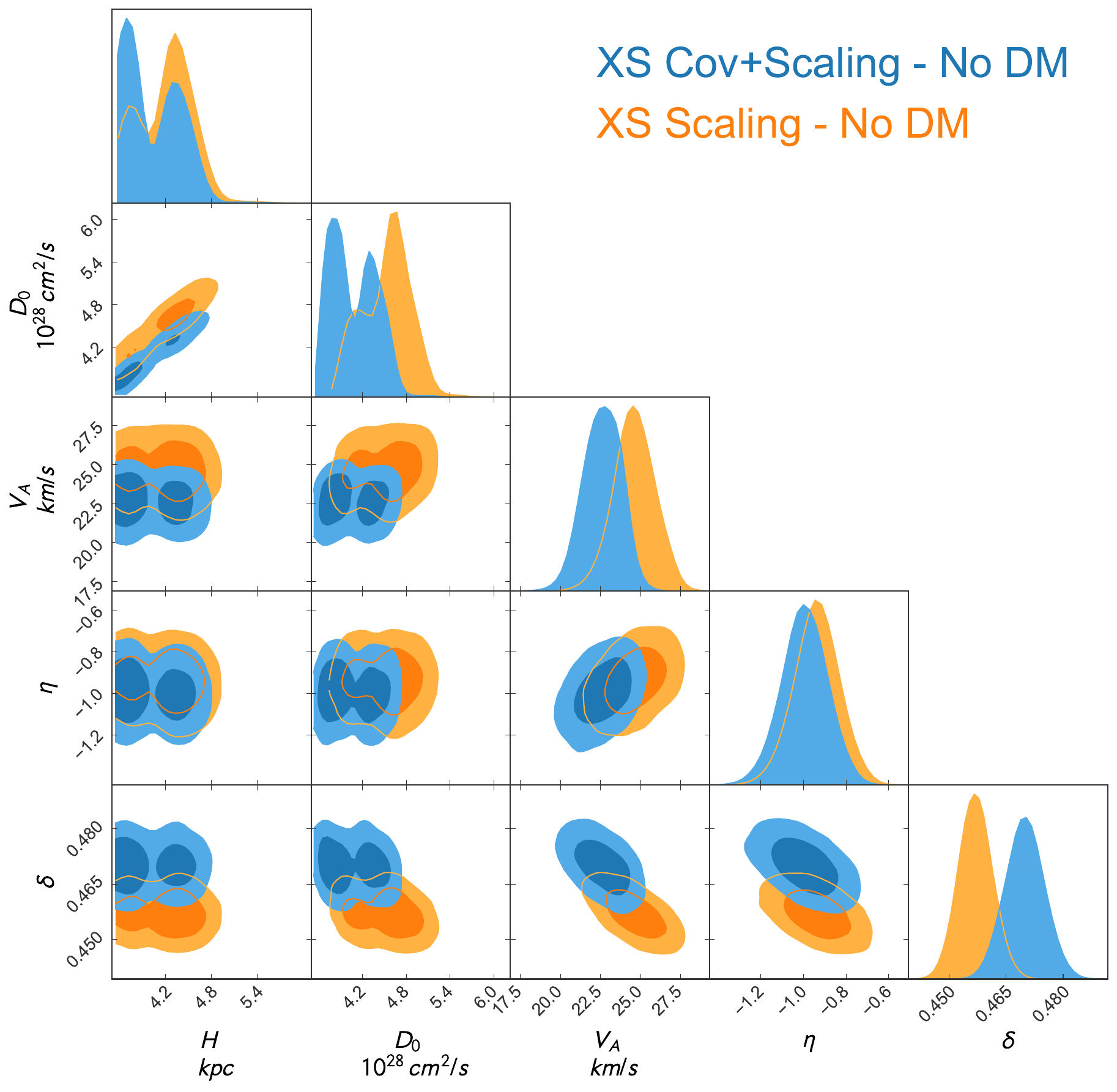}
\includegraphics[width=0.47\textwidth,clip]{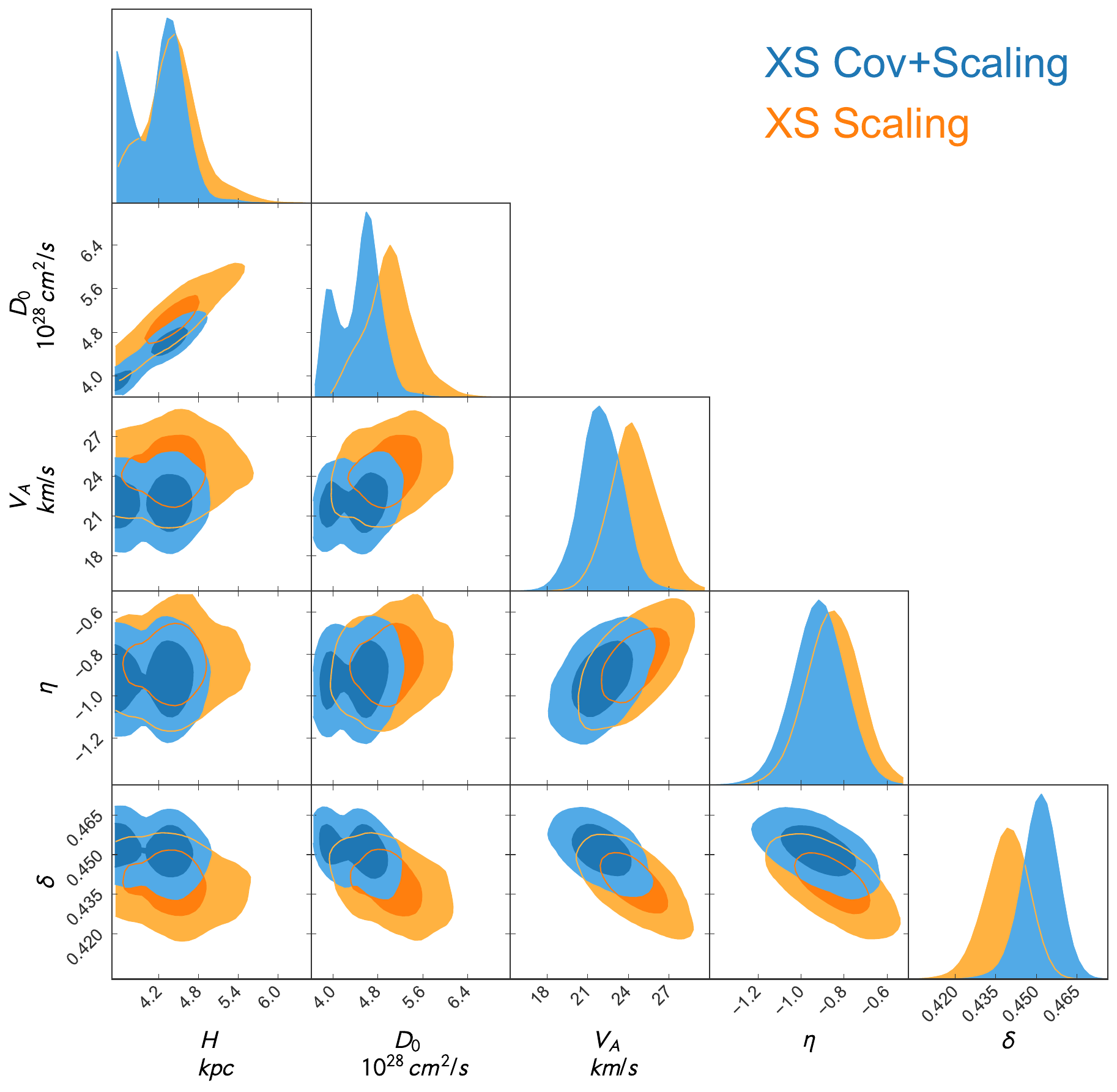} 
\caption{PDFs of the propagation parameters obtained for the analyses performed with correlations in AMS-02 systematic errors. The left panels correspond to the results obtained in the scenario without WIMP contribution, while the right panel show the results for the analyses including this contribution. The contour plots highlight the $68\%$ and $95\%$ credible regions. ``XS Cov $+$ Scaling'' refers to our Canonical analysis and ``XS Scaling'' to the Simplified analysis.}
\label{fig:joint_PDFs}
\end{figure}

\begin{figure}[!b]
\centering
\includegraphics[width=0.37\textwidth,clip]{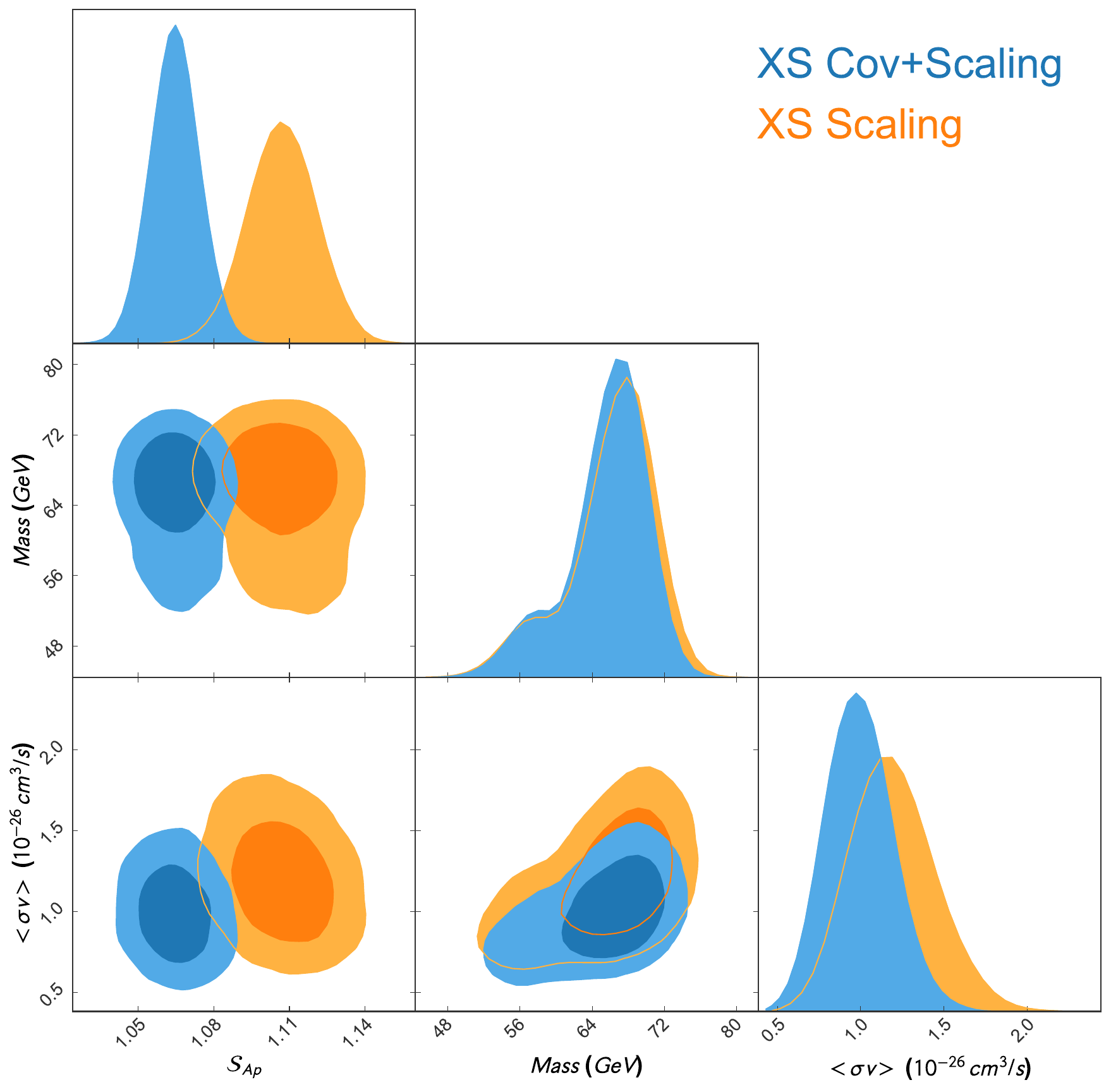} \hspace{1cm}
\includegraphics[width=0.37\textwidth,clip]{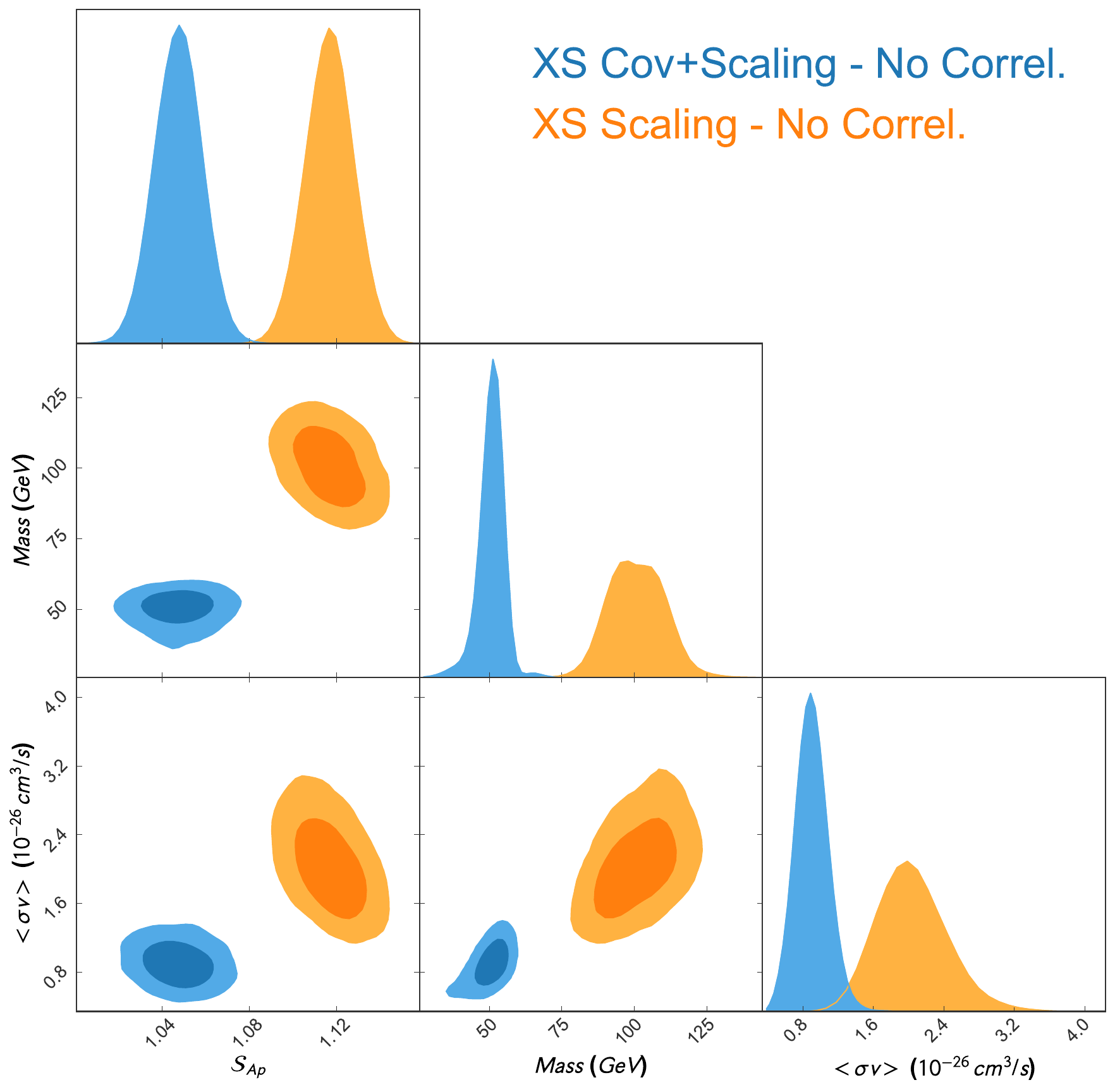} 
\caption{Probability distribution functions for the best-fit antiproton cross section scale, WIMP mass and annihilation rate, $\left< \sigma v \right>$, for the analyses performed including (left panel) and not including (right panel) correlations in the AMS-02 systematic errors. The contour plots highlight the $68\%$ and $95\%$ credible regions. ``XS Cov $+$ Scaling'' refers to our Canonical analysis and ``XS Scaling'' to the Simplified analysis.}
\label{fig:joint_PDFs-DM}
\end{figure}

In this appendix, we show the probability distribution functions (PDFs) of each of the parameters obtained from the analyses explained above. In Figure~\ref{fig:joint_PDFs} we display the PDFs for the different analyses carried out including and not including AMS-02 error correlations and for each scenario considered (with and without contribution from WIMP annihilation). Fig.~\ref{fig:joint_PDFs-DM} displays the PDFs for WIMP mass and annihilation rate, $\left< \sigma v \right>$, for the analyses performed including (right panel) and not including (left panel) correlations in the AMS-02 systematic errors. In the legends of these figures, ``XS Cov $+$ Scaling'' refers to our Canonical analysis and ``XS Scaling'' to the Simplified analysis.

\begin{figure}[h]
\centering
Analyses with only secondary production of $\bar{p}$
\vskip 0.15in
\small{Correlations not included}
\includegraphics[width=0.75\textwidth,clip]{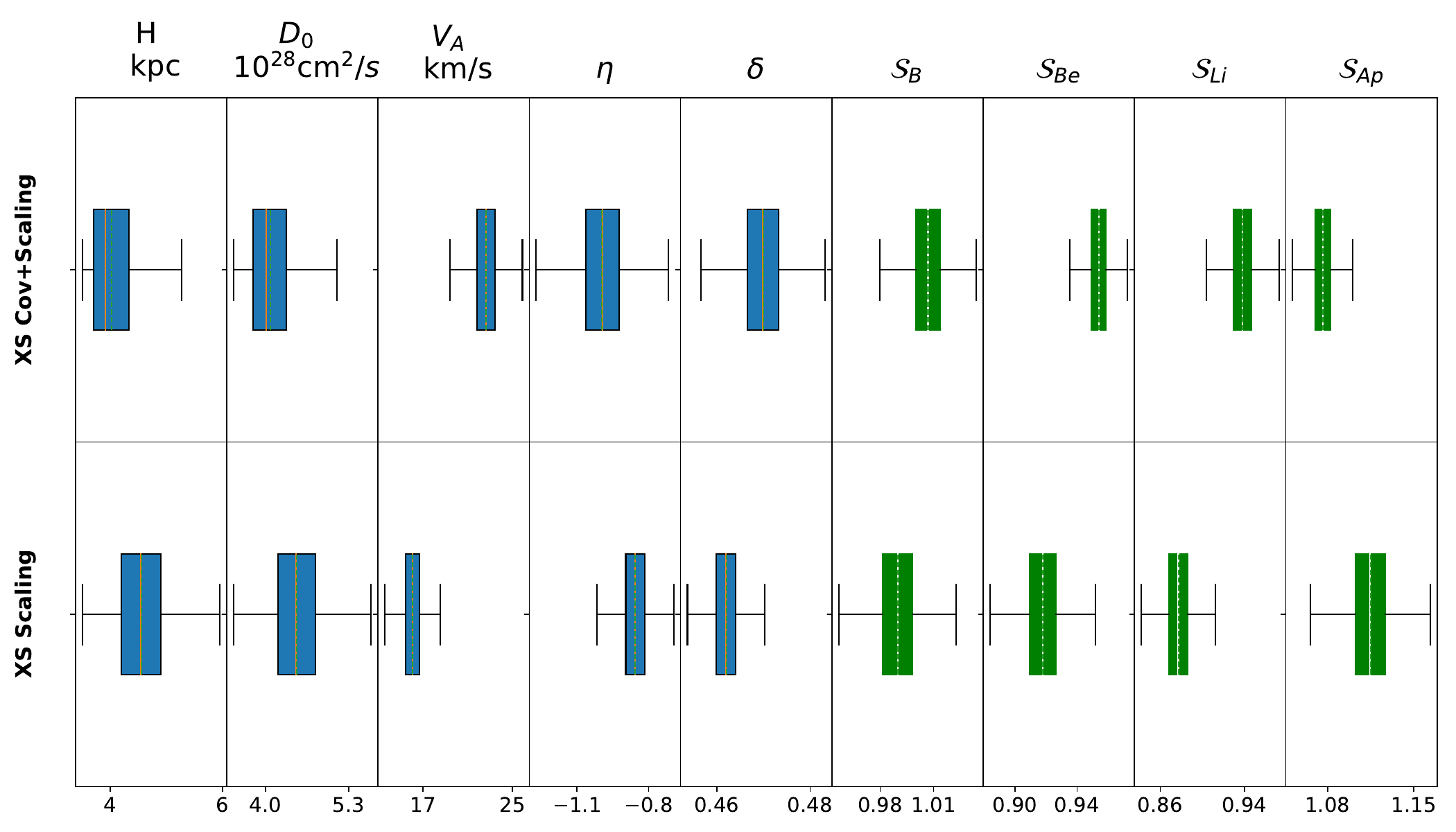} 
\vskip 0.25in

\small{Correlations included}

\includegraphics[width=0.75\textwidth,clip]{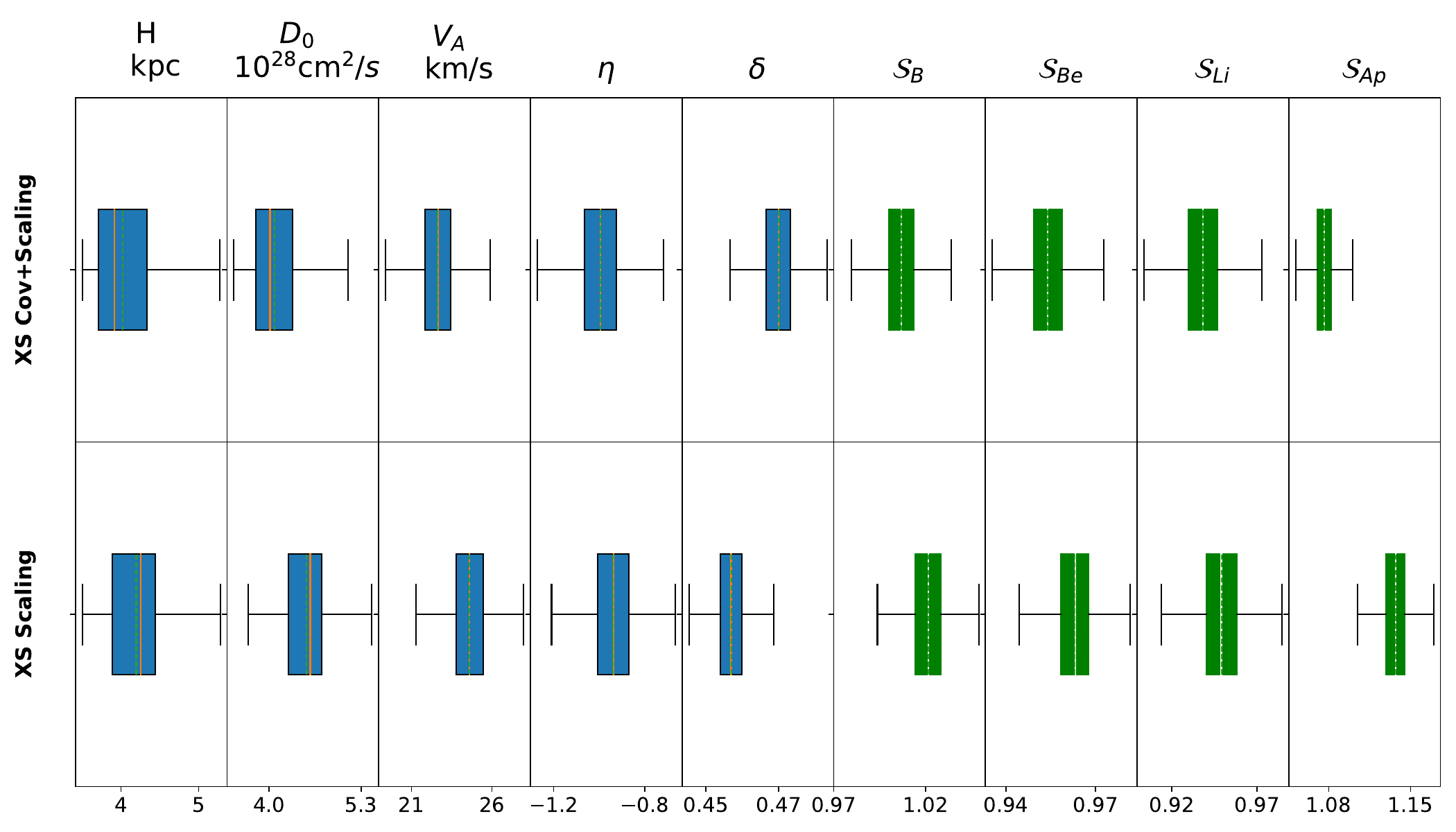} 
\caption{Box-plots describing the distributions for each parameter from the scenario where only secondary production of antiprotons is assumed. In the top panel we show the results obtained from the analysis that no not include correlations in the AMS-02 systematic errors and in the bottom panel those from the analysis including correlations. The median is shown as an orange line inside the boxes, while the mean is represented as a green dashed line. The boxes represent the interquartile region (first and third quartiles of the distribution), while the edges of the whiskers represent the upper and lower extremes of the distribution. As specified above ``XS Cov $+$ Scaling'' refers to our Canonical analysis and ``XS Scaling'' to the Simplified analysis.}
\label{fig:Boxplot_SecAp}
\end{figure}

\begin{figure}[ht]
\centering
Analyses adding WIMP production of $\bar{p}$
\vskip 0.15in
\small{Correlations not included}
\includegraphics[width=0.8\textwidth,clip]{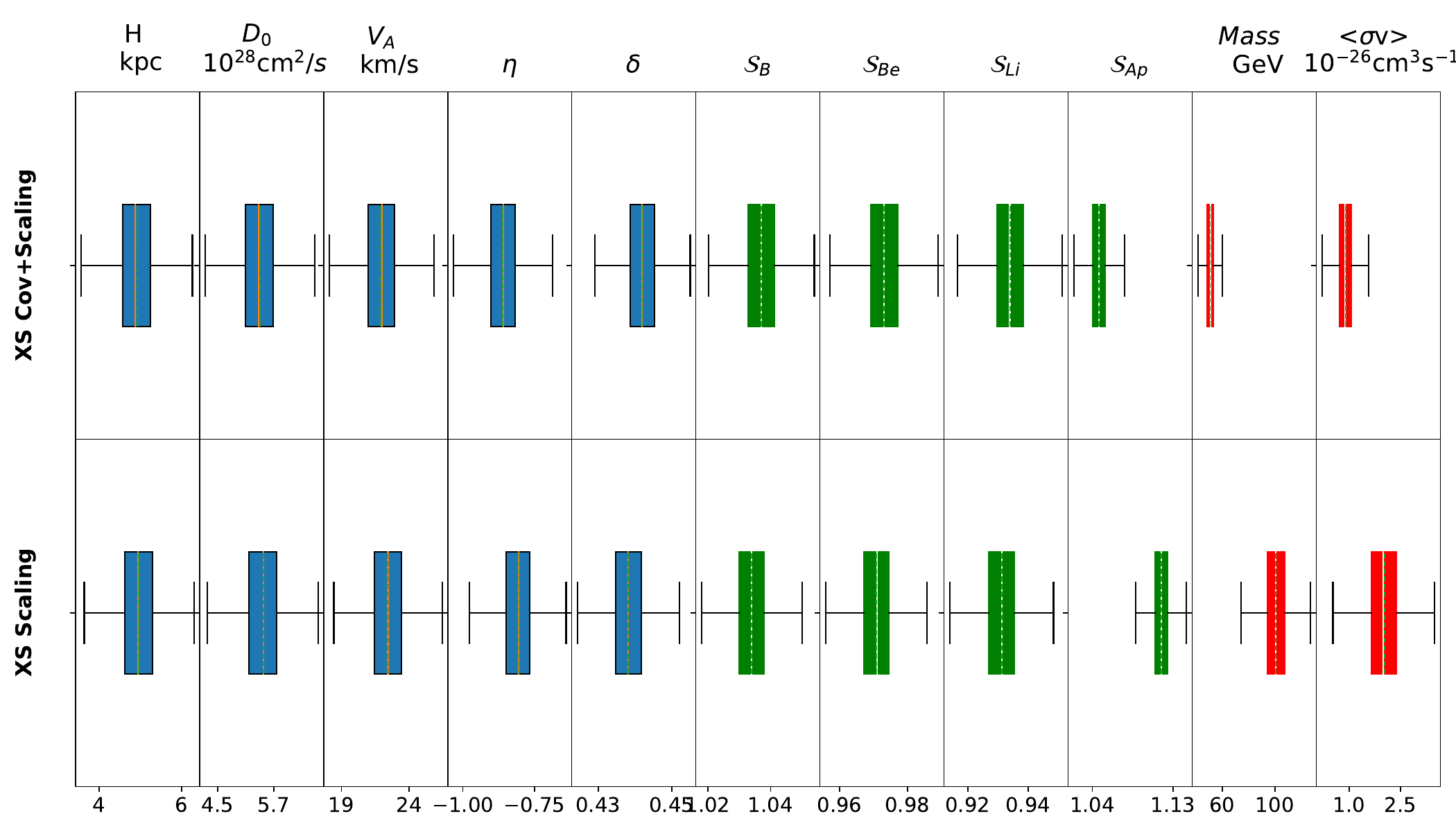} 
\vskip 0.3in
\small{Correlations included}
\includegraphics[width=0.8\textwidth,clip]{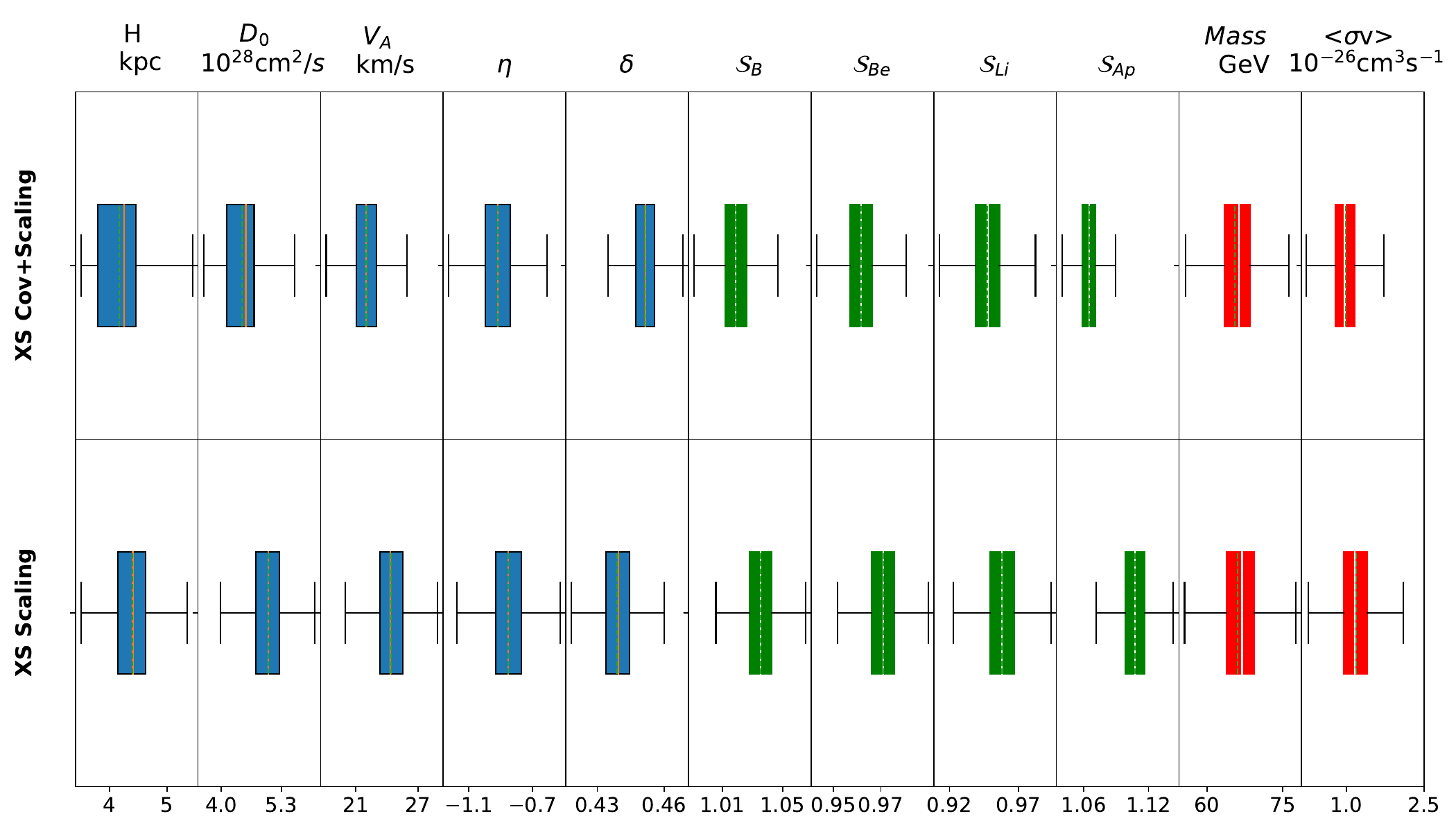} 
\caption{Box-plots describing the distributions for each parameter from the scenario where production of antiprotons from WIMP annihilation is taken into account. In the top panel we show the results obtained from the analysis that no not include correlations in the AMS-02 systematic errors and in the bottom panel those from the analysis including correlations. The median is shown as an orange line inside the boxes, while the mean is represented as a green dashed line. The boxes represent the interquartile region (first and third quartiles of the distribution), while the edges of the whiskers represent the upper and lower extremes of the distribution. As specified above ``XS Cov $+$ Scaling'' refers to our Canonical analysis and ``XS Scaling'' to the Simplified analysis.}
\label{fig:Boxplot_DM}
\end{figure}

In addition, in Figures~\ref{fig:Boxplot_SecAp} and~\ref{fig:Boxplot_DM} we show box-plots that allow us to better understand the uncertainties in the determination of the propagation parameters, scale factors and WIMP annihilation rate and mass. In these plots, the median is shown as an orange line inside the boxes, while the mean is represented as a green dashed line (most of the times, they overlap). The boxes represent the interquartile region (first and third quartiles of the distribution), while the edges of the whiskers represent the upper and lower extremes of the distribution. 

\end{document}